\documentclass[acmtog,nonacm]{acmart}

\usepackage{graphicx}
\usepackage{amsmath}
\usepackage{booktabs}
\usepackage{comment}
\usepackage{amsthm}
\usepackage{algpseudocode} 
\usepackage{color}
\usepackage{multirow}
\usepackage{comment}
\usepackage{overpic}
\usepackage{wrapfig}
\usepackage{hyperref}

\DeclareMathOperator*{\argmin}{arg\,min}

\usepackage[ruled]{algorithm2e} 

\SetAlFnt{\small}
\SetAlCapFnt{\small}
\SetAlCapNameFnt{\small}
\SetAlCapHSkip{0pt}

\begin{document}
\title{Neural Parametric Surfaces for Shape Modeling}

\author{Lei Yang}
\affiliation{%
 \institution{The University of Hong Kong, TransGP}
 \city{Hong Kong SAR}
 \country{China}
}
\email{lyang@cs.hku.hk}

\author{Yongqing Liang}
\affiliation{%
 \institution{Texas A\&M University}
 \city{College station}
 \state{TX}
 \country{USA}
}
\email{lyq@tamu.edu}

\author{Xin Li}
\affiliation{%
 \institution{Texas A\&M University}
 \city{College station}
 \state{TX}
 \country{USA}
}
\email{xinli@tamu.edu}

\author{Congyi Zhang}
\affiliation{%
 \institution{The University of Hong Kong, TransGP}
 \city{Hong Kong SAR}
 \country{China}
}
\email{cyzhang@cs.hku.hk}

\author{Guying Lin}
\affiliation{%
 \institution{The University of Hong Kong}
 \city{Hong Kong SAR}
 \country{China}
}
\email{guyinglin2000@gmail.com}

\author{Alla Sheffer}
\affiliation{%
 \institution{University of British Columbia}
 \city{Vancouver}
 \country{Canada}
}
\email{guyinglin2000@gmail.com}

\author{Scott Schaefer}
\affiliation{%
 \institution{Texas A\&M University}
 \city{College station}
 \state{TX}
 \country{USA}
}
\email{schaefer@cse.tamu.edu}

\author{John Keyser}
\affiliation{%
 \institution{Texas A\&M University}
 \city{College station}
 \state{TX}
 \country{USA}
}
\email{keyser@cse.tamu.edu}

\author{Wenping Wang}
\affiliation{%
 \institution{Texas A\&M University}
 \city{College station}
 \state{TX}
 \country{USA}
}
\email{wenping@tamu.edu}

\begin{abstract}
The recent surge of utilizing deep neural networks for geometric processing and shape modeling has opened up exciting avenues. However, there is a conspicuous lack of research efforts on using powerful neural representations to extend the capabilities of parametric surfaces, which are the prevalent surface representations in product design, CAD/CAM, and computer animation. 
We present \emph{Neural Parametric Surfaces}, the \emph{first} piecewise neural surface representation that allows coarse patch layouts of arbitrary $n$-sided surface patches to model complex surface geometries with high precision, offering greater \emph{flexibility} over traditional parametric surfaces. 
By construction, this new surface representation guarantees \emph{$G^0$ continuity} between adjacent patches and empirically achieves $G^1$ continuity, which cannot be attained by existing neural patch-based methods. The key ingredient of our neural parametric surface is a \emph{learnable} feature complex $\mathcal{C}$ that is embedded in a high-dimensional space $\mathbb{R}^D$ and topologically equivalent to the patch layout of the surface; each face cell of the complex is defined by interpolating feature vectors at its vertices. 
The learned feature complex is mapped by an MLP-encoded function $f:\mathcal{C} \rightarrow \mathcal{S}$ to produce the neural parametric surface $\mathcal{S}$. 
We present a surface fitting algorithm that optimizes the feature complex $\mathcal{C}$ and trains the neural mapping $f$ to reconstruct given target shapes with high accuracy. 
We further show that the proposed representation along with a compact-size neural net can learn a plausible shape space from a shape collection, which can be used for shape interpolation or shape completion from noisy and incomplete input data. 
Extensive experiments show that neural parametric surfaces offer greater modeling capabilities than traditional parametric surfaces.

\end{abstract}

%

%

\ccsdesc[1000]{Computing Methodologies~Parametric curve and surface models}

%
%

\keywords{Deep learning, Surface reconstruction}

\begin{teaserfigure}
    \centering
    \includegraphics[width=0.98\textwidth, trim = 0 520 0 0, clip]{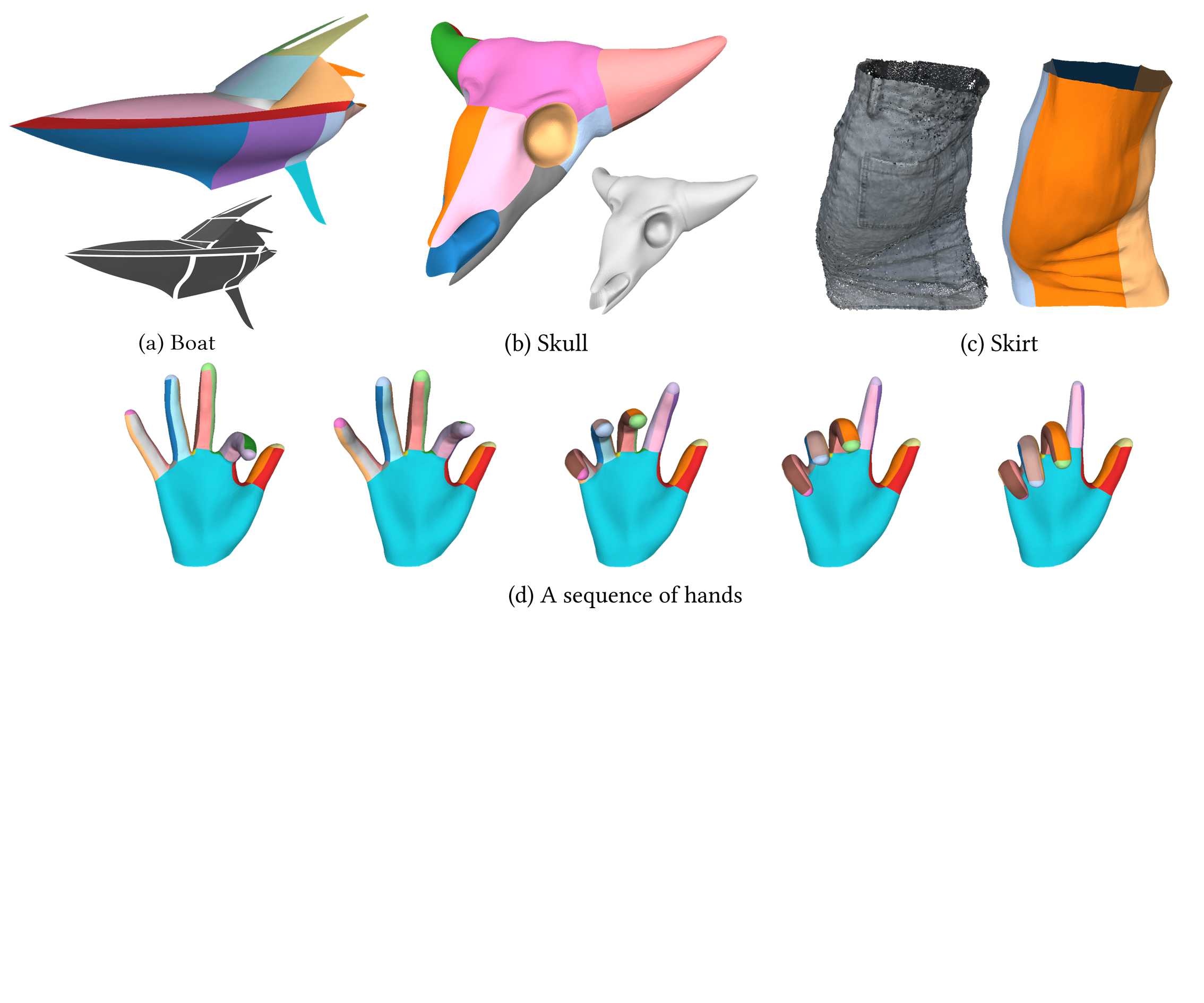}
    \caption{
    We present \textit{Neural Parametric Surfaces}, a novel piecewise parametric representation that extends the capabilities of traditional parametric surfaces to model surface geometry with $n$-sided patches, allowing for the production of surface shapes to align their design semantics (a), for modeling complex surface geometry (b), and for the use of very coarse patch layouts to fit free-form shapes (e.g.\ 4 patches in total in (c)) with high accuracy. This representation can be used to learn a latent shape space from a collection of shapes for different downstream tasks (e.g.\ shape interpolation in (d)).}
    \label{fig:teaser}
\end{teaserfigure}

\maketitle
\section{Introduction}
In this work, we explore the power of neural networks to enhance the capabilities of parametric surfaces, which are the prevalent representation in shape modeling in various forms, such as spline surfaces, subdivision surfaces, Coons patches, etc. Notwithstanding a proliferation of studies in utilizing deep neural networks for geometric processing~\cite{aigerman2022neural,deprelle2022learning,yang2021geometry,morreale2022neural} and shape modeling~\cite{yang2018foldingnet,park2019deepsdf,sitzmann2020implicit,groueix2018papier,guo2022implicit}, there is a conspicuous lack of research efforts on the neural representation of parametric surfaces. Most existing works in this direction are concerned with neural {\em implicit} representations, which represent a shape as the level set of some function over a spatial domain~\cite{takikawa2021neural,martel2021acorn,park2019deepsdf}. Despite their advantages of smoothness and compactness over discrete representations (e.g. point clouds and meshes), the neural implicit representations have difficulty in representing open surfaces, non-manifold surface patches, and surfaces with sharp features commonly present in CAD models, which can be handled naturally by parametric surfaces. 

There have been a few learning-based methods for parametric forms. 
~\cite{groueix2018papier,bednarik2020shape,williams2019deep} adopt a \textit{patch-based} representation to model a given shape as \textit{an atlas of surface patches} overlapping each other. The methods in~\cite{sharma2020parsenet,guo2022complexgen} produce a segmentation from given point cloud data and convert the segmented patches into parametric surfaces. However, none of these works can ensure continuity across adjacent surface patches, which is crucial for applications in product design, reverse engineering, and CAD/CAM.

A major limitation of traditional parametric surfaces (e.g. spline surfaces and subdivision surfaces) is the common assumption of a \emph{rectangular} or \emph{triangular} parametric domain for each surface patch, which precludes the use of semantically meaningful multi-sided surface patches, such as pentagonal or hexagonal patches. Furthermore, since typically lower-degree polynomials (e.g. cubic polynomials) are used as basis functions, the representation power of traditional parametric surfaces is limited in the sense that a large number of refinement patches are necessitated to represent shapes with fine geometric details, especially in shape fitting applications. Therefore, it is desirable to develop more powerful methods to accommodate the modeling of parametric surfaces composed of arbitrary $n$-sided patches that are coarse, semantically meaningful, and smoothly joined.

\begin{figure}
    \centering
    \includegraphics[width=\linewidth, trim=0 200 30 0, clip]{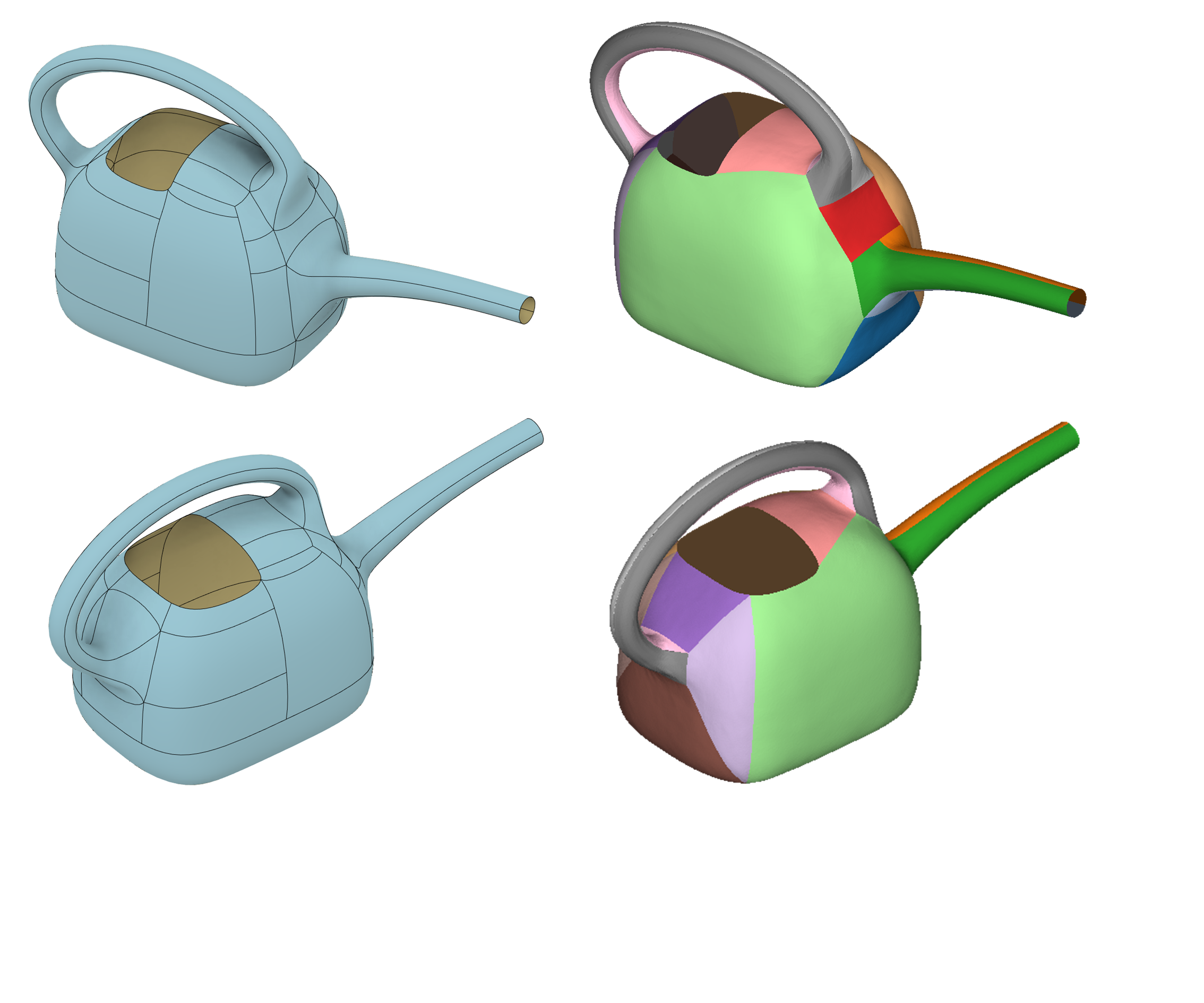}
    \caption{
    To model a \textit{Kettle} surface, the T-spline method (left column) needs 32 quad patches, while our representation (right column) uses fewer (16 here) polygonal patches.}
    \label{fig:comp_tspline}
\end{figure}

We present \textit{Neural Parametric Surfaces}, the \textit{first} piecewise neural surface representation composed of arbitrary $n$-sided surface patches with $G^0$ continuity to model complicated surface geometry, offering greater modeling flexibility than traditional parametric surfaces; see Fig.~\ref{fig:comp_tspline} as an example. The adjacent patches are guaranteed to be continuously joined; furthermore, training with an effective loss function is able to achieve empirically smooth joining ($G^1$ continuity) between the patches. We employ an MLP-encoded function to parameterize each patch so as to allow for coarse patches to model complex shapes due to the increased representation power of individual patches. We demonstrate the applications of this new representation in surface fitting and in learning the shape space of a class of objects for several downstream tasks. All these properties of this new surface representation are attributed to two key components: 1) a \textit{learnable feature complex};  and 2) a \textit{learned continuous mapping function} $f$ that is modeled by deep neural networks.

\textbf{Feature complex}. We introduce a {\em learnable feature complex} $\mathcal{C}$, a 2-complex embedded in higher dimensional space $\mathbb{R}^D$. As a topological structure, $\mathcal{C}$ is defined by a collection of face cells, together with their shared boundary arcs, and vertices. A vertex of ${\mathcal C}$ is represented by a learnable $D$-dimensional vector, called a {\em vertex feature vector}, or {\em vertex feature} for short. 

We adopt the mean value interpolation~\cite{floater2003mean} to define the points on each face of the feature complex from its vertex features. Then an MLP network $f$ is used to map the 2-manifold complex $\mathcal{C}$ from $\mathbb{R}^D$ to $\mathbb{R}^3$ to yield a parametric surface, with each face cell of $\mathcal{C}$ mapped to a surface patch. This allows the \textit{Neural Parametric Surface} to achieve an accurate approximation of the given geometry without the need for refinement schemes used in traditional spline surfaces, thus affording the use of a coarse and semantically meaningful patch layout.

We note that, for each face of the complex, the mean value interpolation induces a linear interpolation on each boundary edge from its two end vertex features. Therefore, all the adjacent faces of the complex have a shared boundary curve, which is a straight line segment given by linear interpolation. Hence, any two adjacent faces of the complex are continuously joined, and so are their corresponding patches of the neural parametric surfaces.

\textbf{Surface fitting}. 
We present a surface fitting method for converting other surface representations (e.g.\ point clouds or polygonal meshes) to the proposed \textit{Neural Parametric Surface}. Given a pre-segmented surface shape,\footnote{We assume the segmentation is either inherent to the shape (as in models using B-Rep representation) or manually prepared by users. The only requirement of this segmentation is that each segment needs to have a disc topology.} we first construct a feature complex $\mathcal{C} \subset \mathbb{R}^D$ of a \textit{Neural Parametric Surface}. This complex is topologically equivalent to the patch layout induced by the partition of the target shape. Then, the surface fitting task is to represent the target shape by a neural parametric surface $\mathcal{S}$ composed of parametric surface patches. Each segment of the target shape is approximated by a patch of $\mathcal{S}$. 

Our fitting method is a learning framework that optimizes the geometry of the feature complex $\mathcal{C}$ and trains the neural mapping function $f$ to produce the desired neural parametric surface. We show the proposed approach can reconstruct, at a high-fidelity level, a diverse set of surface shapes, ranging from CAD models and mechanical parts with smoothly changing surface geometry and sharp features, to garments and organic objects with detailed surface features such as wrinkles. 

\textbf{Leaning from a data collection}. We demonstrate the \textit{Neural Parametric Surface}'s capability of learning from a collection of shapes and optimizing for a shape space of objects. With this capability, the \textit{Neural Parametric Surface} can be used for shape interpolation, a task useful in style design and editing, and can be applied to surface generation from a noisy and incomplete point cloud input.

Our technical contributions are as follows: 
\begin{itemize}
    \item We propose the first piecewise neural parametric surface representation, \textit{Neural Parametric Surface}, that is capable of modeling surfaces with coarse $n$-sided polygonal patches and attaining smooth surface reconstruction of the target geometry, thus greatly extending traditional parametric surfaces.
    \item A learnable feature complex is introduced. The feature complex is defined in a high-dimensional embedding space and decoded with a neural mapping function into a parametric surface with \textit{smooth joining} between the resulting patches.
    \item An efficient surface fitting algorithm is presented for converting other surface representations to a neural parametric surface. 
    \item We demonstrate neural parametric surfaces can be used to learn the shape space of a shape collection for performing shape interpolation and shape generation from noisy and/or incomplete point clouds. 
\end{itemize}

\section{Related works}\label{sec:related_works}

\textbf{Neural implicit representations. } 
Several studies~\cite{chibane2020neural,sitzmann2020implicit,gropp2020implicit} have exploited the approximation capabilities of deep neural networks to enhance implicit representations that define a target surface geometry as the zero-level set of a scalar function. These works have enabled learning a shape space from a collection of data~\cite{park2019deepsdf,deng2021deformed} or using regular feature grids to enhance the efficacy of ordinary MLP networks~\cite{martel2021acorn,takikawa2021neural}. To model complex geometry, several recent works~\cite{tretschk2020patchnets,zhang2022implicit} adopt a compositional approach, which represents each part of a given 3D shape as the zero-level set of a learned implicit field and blends multiple such fields corresponding to different parts to fit the given 3D shape. 

Due to the continuous nature of a neural network, it is challenging to represent sharp features present in CAD models using neural implicit representations. A recent study~\cite{guo2022implicit} proposed to represent the surface patches as neural half-spaces and then compose them into a watertight model via Boolean operations as used in Constructive Solid Geometry. This can be cumbersome as the network must additionally take care of the extended parts of the zero-level sets that approximate the surface patches; otherwise, artifacts may be observed. 

In contrast to neural {\em implicit surface} representations, we propose a new neural {\em parametric surface} representation and show that it inherits many merits of traditional parametric surface representations, such as modeling open surfaces, sharp features, and non-manifold surface patches that neural implicit representations now struggle to achieve. Similar to traditional parametric representations, our neural parametric surfaces are easy to render as compared to {\em implicit surface} representations which require a Marching Cubes-like algorithm to contour the zero-level set.

\begin{figure*}[t!]
    \centering
    \begin{overpic}
    [width=\linewidth, trim=0 850 0 0, clip]{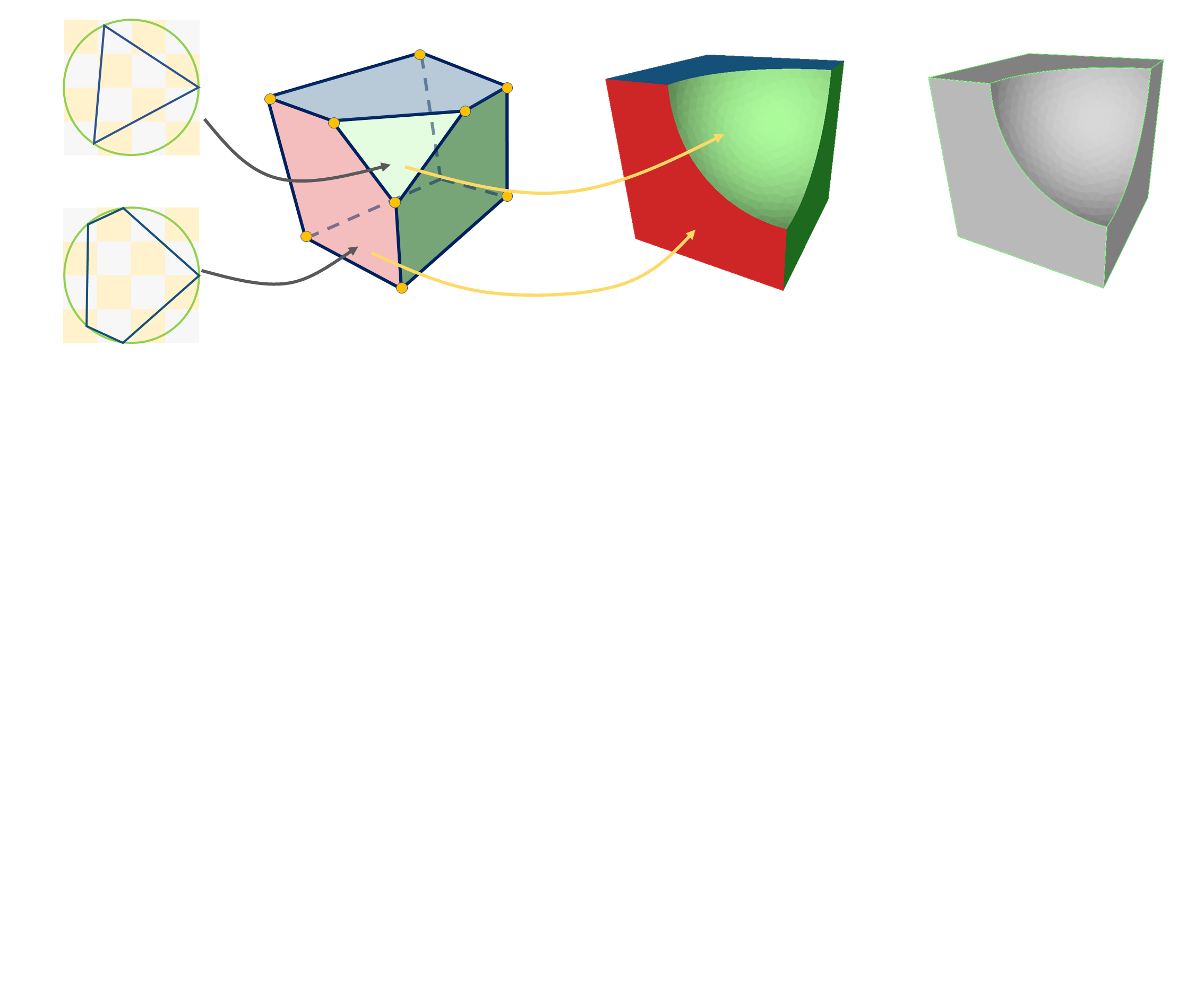}
    \put(44,10){$f_{\theta}$}
    \put(23,11){$g$}
    \put(23,27){$\mathbf{z}_k$}
    \put(32,21){$C_i$}
    \put(60,23){$S_i$}
    \put(30,14){$C_j$}
    \put(57,15){$S_j$}
    \put(10,25){ $\Omega_i$}
    \put(10,10){ $\Omega_j$}
    \put(2,1){(a) Polygonal domain}
    \put(25,1){(b) Feature complex $\mathcal{C}$}
    \put(50,1){(c) Neural parametric surface $\mathcal{S}$}
    \put(80,1){(d) Target surface $\mathcal{M}$}
    \end{overpic}
    \caption{\textbf{Design for Neural Parametric Surfaces.} 
    For a given target surface $\mathcal{M}$, we construct a feature complex $\mathcal{C}$ based on a set of learnable features $\{\mathbf{z}_k\}$ defined at complex cells' vertices (yellow nodes). Each complex face ($C_i$) is parameterized by a planar $n$-sided polygonal domain $\Omega_i$ using the mean value interpolation $g$. 
    We train the complex embedding $\{\mathbf{z}_k\}$ and the mapping function $f_{\theta}$ to produce a neural parametric surface $\mathcal{S}$ that approximates $\mathcal{M}$ with high fidelity and preserves the partitioning (each patch delineated by green boundary curves) it carries. 
    }
    \label{fig:feature_complex}
\end{figure*}
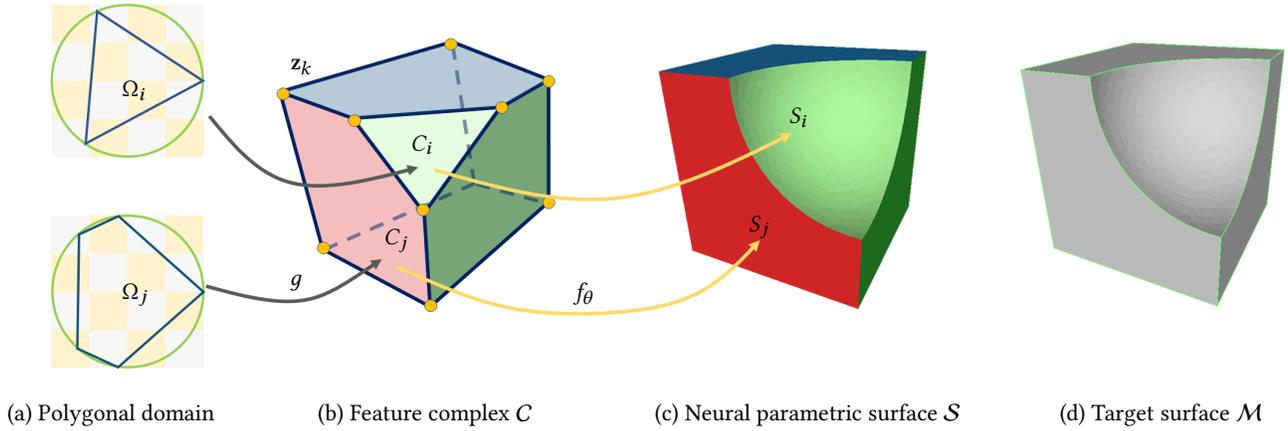

\textbf{Neural patch-based representations. }
Some pioneering works have been proposed to represent a 3D shape either as a single patch~\cite{yang2018foldingnet} or multiple patches, as in AtlasNet~\cite{groueix2018papier}. Similar to traditional parametric representations, AtlasNet defines these patches on a 2D rectangular domain and lifts them to 3D using multiple learnable mapping functions to approximate a target shape. 
However, for complex 3D surface shapes (as shown in some examples in Fig.~\ref{fig:teaser}), using solely these rectangular patches can result in surface reconstructions with holes. AtlasNet, hence, circumvents this (visually) by wrapping the generated patches around the target shape to produce the surface reconstruction. However, these surface patches actually overlap with their surrounding patches rather than seamlessly stitching with each other along their boundaries, which is not desirable for downstream applications like product design and CAD/CAM. 

A number of studies~\cite{williams2019deep,groueix20183d,deprelle2019learning,gadelha2021deep,low2022minimal} have built upon AtlasNet's framework and extended it in various aspects, for example, further enhancing its ability in modeling shape collections~\cite{groueix20183d,deprelle2019learning}, mitigating the issue of overlapping patches~\cite{gadelha2021deep}, and enabling the capability of handling holes within an individual parametric patch~\cite{low2022minimal}. 
However, like AtlasNet, these approaches also assume that each chart is \textit{ topologically rectangular}. They also use \textit{distinct} mapping functions for each chart. As a result, adjacent 3D surface patches still overlap with each other in the final reconstruction. To produce a seamless representation, \cite{williams2019deep} requires an extra step to convert its output into an implicit field using Poisson reconstruction. 

In contrast to AtlasNet and its variants, which separately map 2D parametric domains to individual 3D surface patches, our approach employs an optimized feature complex as an intermediate geometric entity to stitch multiple 2D domains in a learnable feature space. We then train a shared mapping function to project each point from this feature complex to 3D space. Consequently, our approach ensures that the resulting geometry is piecewise smooth and that adjacent surface patches connect continuously at their shared boundaries. While our formulation requires a given patch layout for the target shape as input, this layout can be easily generated using off-the-shelf tools like~\cite{livesu2020loopycuts}. As shown in our experiment, it is difficult for AtlasNet-based approaches (e.g.,~\cite{deng2020better}) to accommodate patch layouts containing $n$-sided polygonal patches due to their dependency on rectangular parametric domains.

In summary, the advantages of our proposed representation are as follows: 1) our \textit{Neural Parametric Surface} is a continuous representation, i.e.\ no gaps between two adjacent patches; 2) its surface patches can be arbitrary $n$-sided, rather than just rectangular as in traditional parametric representations or recent learning-based representations; and 3) meshing and rendering of the resulting shape are straightforward as will be described later, without the need for contouring the surface geometry required by neural implicit representations.

\textbf{Learning parameterization for 3D shapes.} 
Several works propose to learn a parameterization $g$ that projects points from a given 3D shape onto a 2D parametric plane~\cite{morreale2021neural,deprelle2022learning} or multiple 2D charts~\cite{pokhariya2022discretization}. However, since parameterizing topologically non-trivial surfaces on a rectangle or disk domain inevitably introduces cut seam(s), surfaces reconstructed using the inverse parameterization $g^{-1}$ will carry the cut seams. Therefore, a simple planar parameterization such as \cite{deprelle2022learning} does not ensure seamless reconstruction results. In contrast, our approach can preserve the desired continuity between adjacent patches in the reconstructed results by leveraging the feature complex that captures the topological structure of the target shape. 

\textbf{Constructing a patch layout from 3D shapes. }
There is a line of work to predict the segmentation of a 3D geometry and reconstruct the shape using parametric primitives~\cite{yu2022piecewise}, Coons patches~\cite{smirnov2020learning}, B-spline surfaces~\cite{sharma2020parsenet}, or a neural parametric patch defined on a rectangle domain~\cite{guo2022complexgen}. Our work is complementary to these studies. For example, the patch layout predicted by~\cite{smirnov2020learning} can be used as an input to our approach, while the representational power of our Neural Parametric Surface over Coons patches can be leveraged to represent more complex geometry (cf.\ Fig.~\ref{fig:comp_w_coons}). 

Many previous methods are proposed to produce a (quad) patch layout for geometry processing~\cite{cohen2004variational,tarini2004polycube,livesu2020loopycuts,born2021layout,campen2017partitioning}. We show that our Neural Parametric Surface can take as input the $n$-sided segmented patches produced by~\cite{livesu2020loopycuts,cohen2004variational} and convert a mesh-based representation to a neural piecewise parametric surface.

\section{Neural Parametric Surfaces}\label{sec:methodology}

\subsection{Constructing Feature Complex}
\label{sec:neural_complex}

Given a 3D surface geometry along with an associated patch layout (as shown in Fig.~\ref{fig:feature_complex}(d)), we introduce a 2D feature complex ${\mathcal C}$ (see Fig.~\ref{fig:feature_complex}(b)) that is embedded in a high-dimensional feature space $\mathbb{R}^D$ ($D=128$ in this work). This feature complex maintains the same topological structure as the patch layout. 
The embedding of feature complex ${\mathcal C}$ is determined by its vertices, each represented as a  $D$-dimensional vector, referred to as {\em vertex features}. Each $n$-sided face cell of ${\mathcal C}$ is obtained via mean value interpolation from the vertices on an $n$-sided planar polygon. Then, a \textit{shared} MLP-encoded function $f: \mathbb{R}^D \rightarrow \mathbb{R}^3$ is employed to map the 2-complex ${\mathcal C}$ onto a continuous, piecewise smooth surface ${\mathcal S} \in \mathbb{R}^3$, in which each patch corresponds one-to-one with the face cells of ${\mathcal C}$, as illustrated in Fig.~\ref{fig:feature_complex}(c). The vertex features are learned during a surface fitting procedure. This neural parametric surface representation can be flexibly defined to depict various shapes, based on either user-designed or automatically computed patch layouts. 

Consider a face cell ${\mathcal C}_i$ within the complex ${\mathcal C}$. On one hand, ${\mathcal C}_i$ is mapped to a surface patch ${\mathcal S}_i \subset \mathbb{R}^3$ via an MLP-encoded function $f_\theta:  \mathbb{R}^D \to \mathbb{R}^3$. On the other, the face cell ${\mathcal C}_i$ is defined by the mean-value interpolation of its vertex features, which parameterizes ${\mathcal C}_i$ over a predefined 2D $n$-sided polygon $\Omega_i$ through the function $g: \mathbb{R}^2 \to \mathbb{R}^D$. Consequently, the parameterization of the surface patch ${\mathcal S}_i$ is defined by the composite function $h \equiv g\circ f_\theta: \mathbb{R}^2 \to \mathbb{R}^3$. Collectively, the union of these surface patches ${\mathcal S}_i$ forms the neural parametric surface. The learnable parameters of this surface representation consist of the vertex features of complex ${\mathcal C}$ and the weights of the shared MLP $f_\theta$. 

\begin{figure}[h!]
    \centering
    \begin{overpic}
    [width=0.7\linewidth, trim=100 850 500 0, clip]{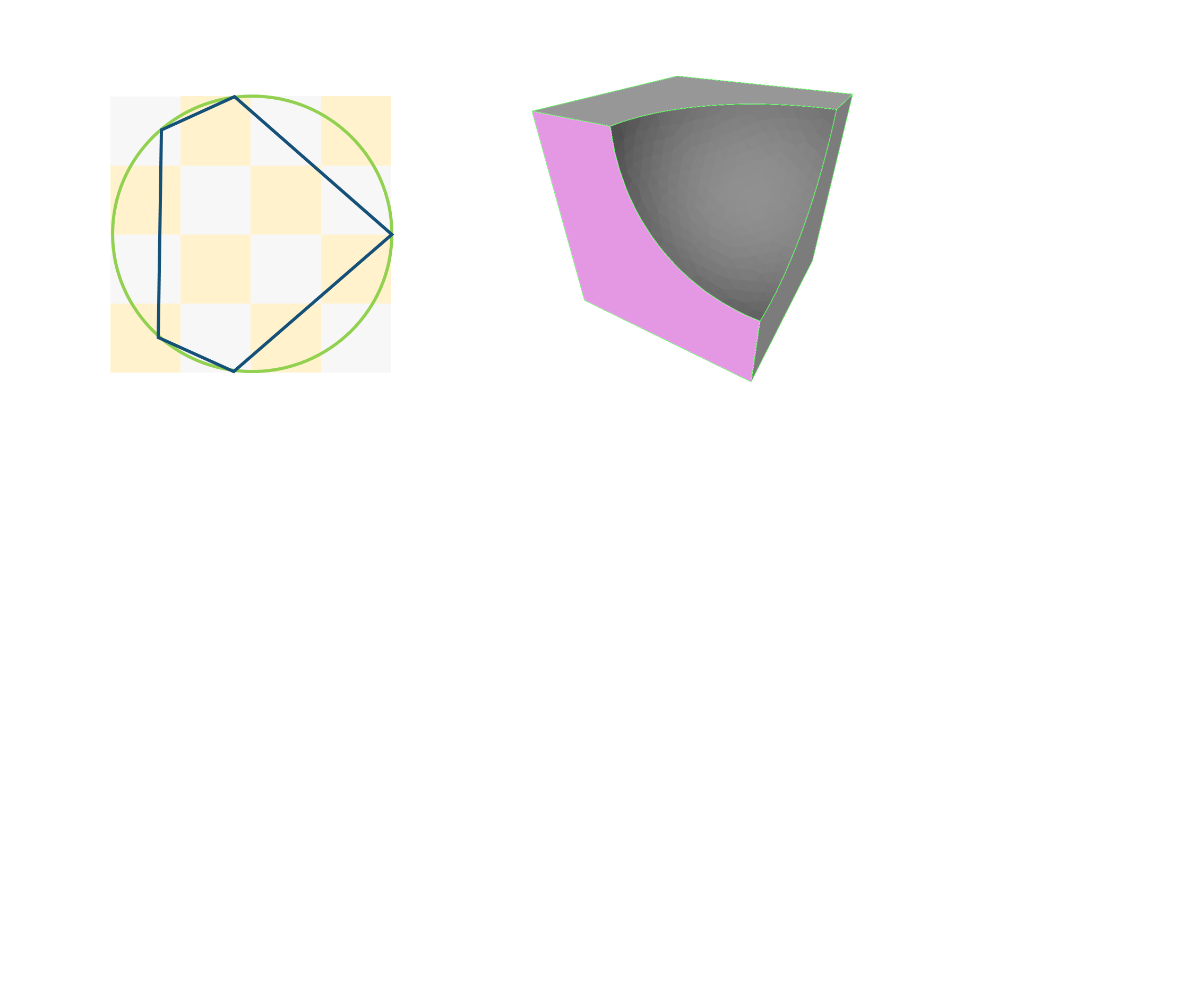}  
    \put(42,22){$\mathbf{u}_0$}
    \put(23,1){$\mathbf{u}_4$}
    \put(23,41){$\mathbf{u}_1$}
    \put(6,36){$\mathbf{u}_2$}
    \put(5,7){$\mathbf{u}_3$}
    \put(60,11){$\mathbf{p}_{0}$}
    \put(84,0){$\mathbf{p}_{1}$}
    \put(54,36){$\mathbf{p}_{4}$}
    \put(64,32){$\mathbf{p}_{3}$}
    \put(82,9){$\mathbf{p}_{2}$}
    \end{overpic}
    \caption{The pink polygonal surface patch (right) is mapped to a pentagon on the 2D plane (left). The vertices of this pentagon sequentially correspond to those of the surface patch.}
    \label{fig:parametric_domain}
\end{figure}

For an $n$-sided patch ${\mathcal S}_i$, its \textit{parametric domain} $\Omega_i \subset \mathbb{R}^2$ is constructed to be an $n$-sided convex polygon inscribed to a unit circle centered at the origin, with its vertices $\mathbf{u}_j$ in correspondence with the corner vertices $\mathbf{p}_j$ of the patch ${\mathcal S}_i$. The 2D corners $\mathbf{u}_j$, $j=1,2,\ldots, n$ are placed on the circle so that the arc length between $\overline{\mathbf{u}_j\mathbf{u}_{j+1}}$ over the circle circumference is proportional to the boundary lengths of $\overline{\mathbf{p}_j\mathbf{p}_{j+1}}$. See Fig.~\ref{fig:parametric_domain} as an example.

Then, the parameterization, $g$, of a face cell ${\mathcal C}_i \subset \mathbb{R}^D$ over $\Omega_i \subset \mathbb{R}^2$ is given by the following interpolation based on mean value coordinates:

\begin{equation}\label{eq:mean_value_interpolation}
    \mathbf{z}(\mathbf{u}) = g(\mathbf{u}) = \sum_{j=1}^n\lambda_j(\mathbf{u})\mathbf{z}_{j},
\end{equation}
where $\mathbf{z}_{j}$ are the vertex feature vectors of the cell ${\mathcal C}_i$, and $\lambda_j(\mathbf{u})$ are the mean value coordinates of the parameter value $\mathbf{u} \in \Omega_i$ associated with the corresponding vertices $\mathbf{u}_j$ of the polygon $\Omega_i$.

\begin{figure*}[ht!]
    \centering
    \begin{overpic}
    [width=\linewidth, trim=0 650 250 0, clip]{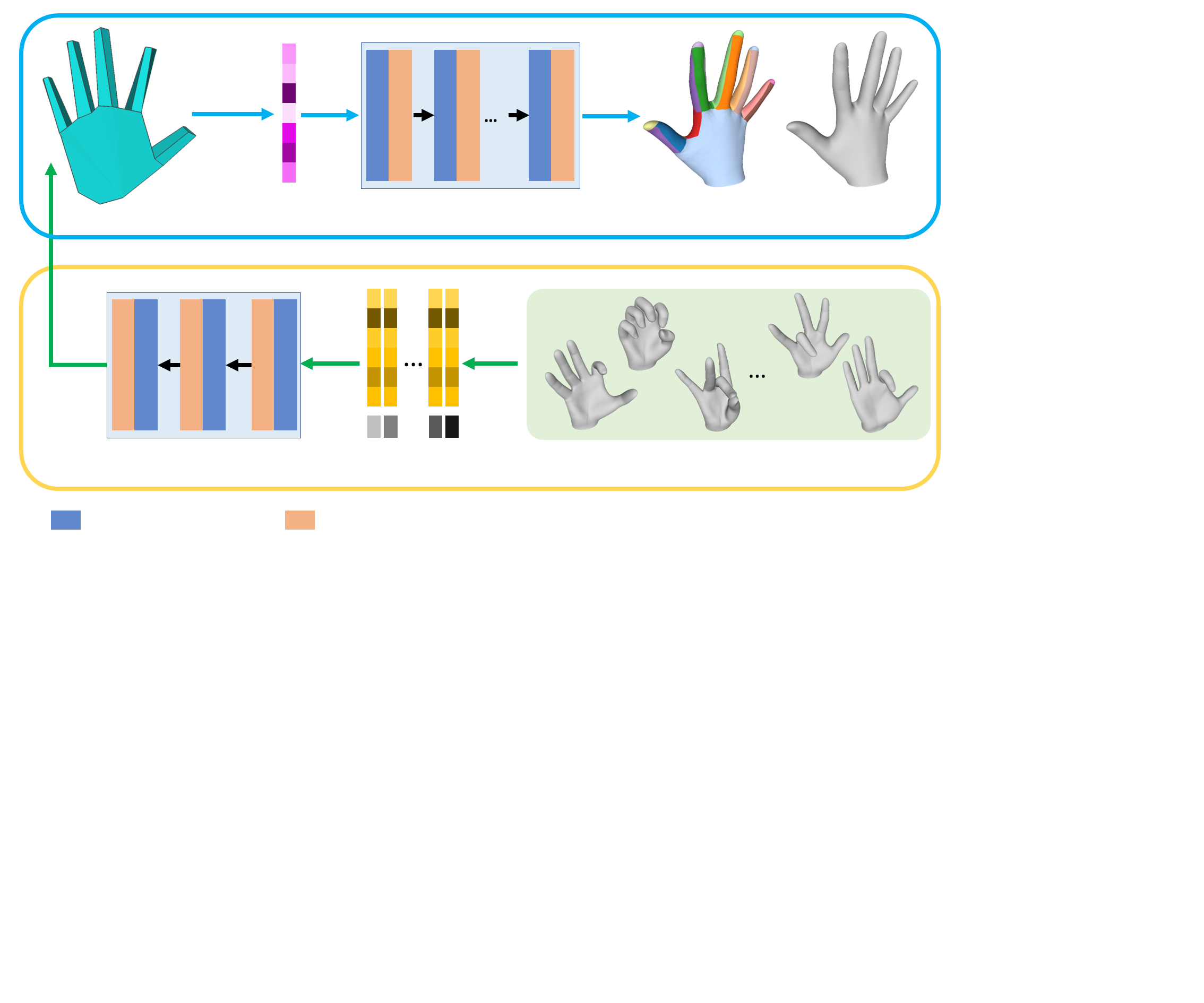}  
    \put(10,31){Feature Complex $\mathcal{C}$}
    \put(26,33){$\mathbf{z} \in \mathbb{R}^D$}
    \put(38,31){MLP-Encoded Map $f$}
    \put(63,31){Output Shape $\mathcal{S}$}
    \put(77,31){Target Shape $\mathcal{M}$}
    \put(36,8){$\mathrm{cat}(\mathbf{c}_m, P_k)$}
    \put(13,7){Broadcast Decoder $h$}
    \put(66,7){Training Dataset}
    \put(9,1.5){Fully-Connected Layer}
    \put(32,1.5){SoftPlus Activation}
    \put(94,40){(a)}
    \put(94,16){(b)}
    \end{overpic}
    \caption{\textbf{Overview of the training pipeline for Neural Parametric Surfaces.}
    (a) Geometric decoder: The top row shows our shape fitting pipeline that jointly optimizes the feature complex $\mathcal{C}$ and MLP-encoded mapping function $f$ to model a single shape with the Neural Parametric Surface representation.  
    (b) Broadcast decoder: To learn from a shape collection, we extend the geometric decoding pipeline with a broadcast decoder $h$ which learns a shape space for synthesizing a shape-specific complex $\mathcal{C}_m$. The input to a broadcast decoder is the broadcast concatenation of a shape code $\mathbf{c}_m$ (orange) and different vertex positional tokens $P_k$ (each $P_k$ corresponding to one complex vertex).
    The entire pipeline of (a) and (b) can learn a latent shape space that encodes a shape collection for various tasks, such as latent shape interpolation or shape editing.
    }
    \label{fig:pipeline}
\end{figure*}
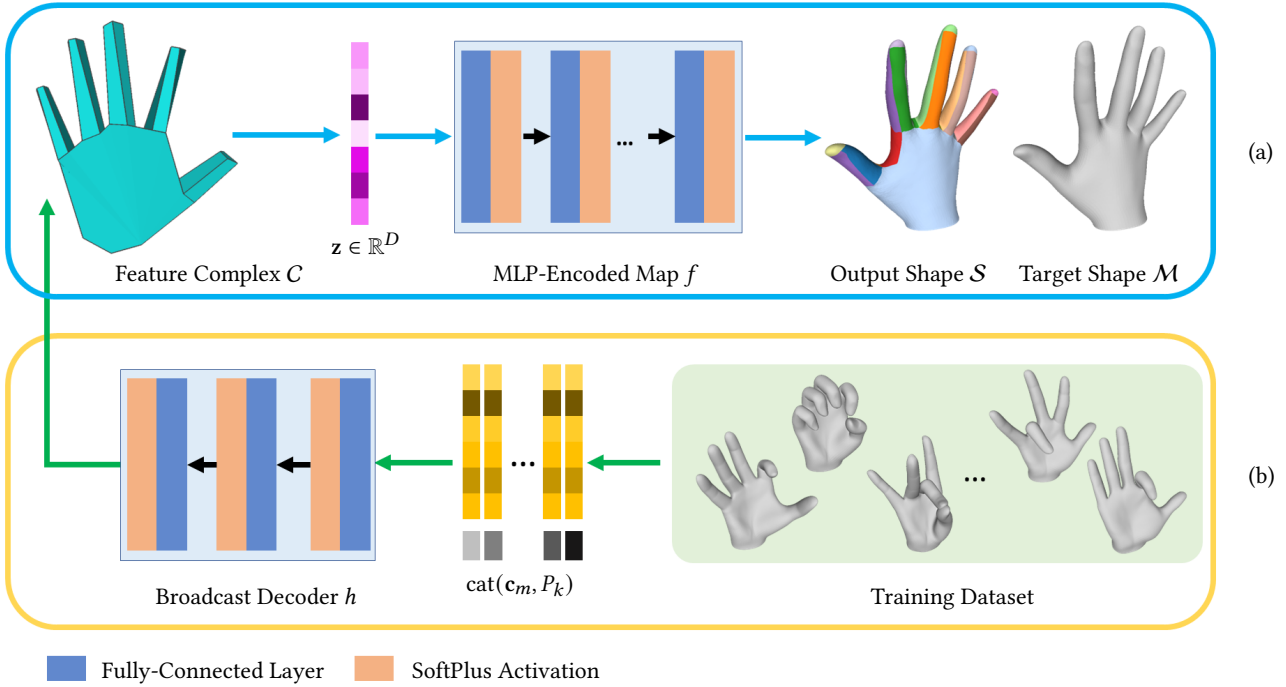

Now consider two adjacent faces ${\mathcal C}_i$ and ${\mathcal C}_j$. 
Because mean value interpolation reduces to linear interpolation along the boundary edges of a face, the edge $\partial {\mathcal C}_{i,j}$ shared by ${\mathcal C}_i$ and ${\mathcal C}_j$ receives the same embedding in the feature space, based on vertex features defined at the edge's endpoints. Consequently, the constructed feature complex $\mathcal{C}$ is \textit{continuous everywhere}. Together with the continuous function $f_\theta$, all feature vectors along $\partial {\mathcal C}_{i,j}$ of two faces ${\mathcal C}_i$ and ${\mathcal C}_j$ are mapped to a 3D curve shared by the corresponding surface patches ${\mathcal S}_i$ and ${\mathcal S}_j$. 
This ensures the continuity of the resulting piecewise parametric surface representation. 
The assurance of continuity between adjacent patches distinguishes our Neural Parametric Surface representation from AtlasNet-like representations.

\subsection{Optimizing Neural Parametric Surfaces}
\label{sec:training}

To derive a \textit{Neural Parametric Surface} fitting a given surface shape, we jointly optimize (1) the feature vectors $\mathcal{Z} = \{\mathbf{z}_k\}$ that define the geometry of $\mathcal{C}$ and (2) the parameters $\theta$ of $f_\theta$. This training pipeline is illustrated in Fig.~\ref{fig:pipeline}(a). Later, we show how this training pipeline can be extended to learn a latent morphable space of a shape category for a diverse set of downstream tasks; see Fig.~\ref{fig:pipeline}(b).

\textbf{Surface fitting.}
We first elaborate on the procedure for fitting our \textit{Neural Parametric Surface} to a single shape. 
The input data for our training pipeline consists of a set of points, each accompanied by its corresponding unit normal vector and a label that indicates the patch to which it belongs. We will discuss later how to generate a patch layout (Sec.~\ref{sec:implementation}) 

Specifically, to reconstruct $\mathcal{M}=\{\mathcal{M}_i\}$ using surface $\mathcal{S}=\{\mathcal{S}_i\}$, we minimize the following loss objective: 
\begin{equation}\label{eq:reconstruction_loss}
    \begin{split}
    L_{\mathrm{recon}} &= 
    L_{\mathrm{anchor}} + \lambda_{\mathrm{surface}}L_{\mathrm{surface}} + \lambda_{\mathrm{normal}}L_{\mathrm{normal}}\\ 
    &+ \lambda_{\mathrm{smooth}}L_{\mathrm{smooth}} + \lambda_{\mathrm{fair}}L_{\mathrm{fair}} \\
    &+ \lambda_{\mathrm{uniform}}L_{\mathrm{uniform}} +\lambda_{\mathrm{aspect}}L_{\mathrm{aspect}}.
\end{split}
\end{equation}

The \emph{anchor deviation} term $L_{\mathrm{anchor}}$ measures the deviation from the corner vertices $\mathbf{x}_k = f_\theta(\mathbf{z}_k)$ of neural parametric patches ${\mathcal S}_i$ to the corresponding corner vertices $\mathbf{p}_k \in M_i$, 
\begin{equation}\label{eq:node_fitting}
    L_{\mathrm{anchor}} = \sum_{k} \|\mathbf{x}_k - \mathbf{p}_k\|_2.
\end{equation}
The anchor term provides a one-to-one correspondence that is important at the early stage of the optimization when the \textit{Neural Parametric Surface} approximation is distant from the target geometry.

The \emph{surface fitting} term $L_{\mathrm{surface}}$ fits a neural parametric patch $\mathcal{S}_i$ to points sampled from $\mathcal{M}_i$. 
We denote $\mathbf{x}_j = f_{\theta}(\mathbf{z}_j)$ as the points sampled from a neural parametric patch $\mathcal{S}_i$. 
Specifically, we compute for each sample point $\mathbf{x}_j$ the closest point from the target patch $\mathcal{M}_i$ using the Euclidean distance metric. Thus, a set of paired points is obtained. Likewise, we obtain another set of paired points by computing for each $\mathbf{p}_j \in \mathcal{M}_i$ its closest point on $\mathcal{S}_i$. 
We denote the two sets of paired points from all patches as $\{(\mathbf{x}_j, \mathbf{p}_j)\}$. 
Therefore, we have 
\begin{equation}\label{eq:surface_fitting}
    L_{\mathrm{surface}} =
    \sum_{\{(\mathbf{x}_j, \mathbf{p}_j)\}}
    \|\mathbf{x}_j - \mathbf{p}_j\|_2 + \beta
    \|{\mathbf{n}_j}^T(\mathbf{x}_j - \mathbf{p}_j)\|_2.
\end{equation}
Here, $\beta$ is a weighting coefficient and is set to $0.1$ throughout the optimization procedure, and $\mathbf{n}_j$ is the unit normal vector at sample point $\mathbf{p}_j$.

We use the term $L_{\mathrm{normal}}$ to enforce the normal consistency between each pair of the corresponding points $\{(\mathbf{x}_j, \mathbf{p}_j)\}$:
\begin{equation}
\label{eq:normal}
    L_{\mathrm{normal}} = 
    \sum_{\{(\mathbf{x}_j, \mathbf{p}_j)\}}
    (1 - \mathbf{n}_j^T \mathbf{n}(\mathbf{x}_j) ),
\end{equation}
where $\mathbf{n}(\mathbf{x}_j)$ is the unit normal vector at $\mathbf{x}_j$ that can be derived analytically as follows. We first compute the partial derivative $\partial \mathbf{x}_j / \partial \mathbf{u}$, exploiting the differentiability of the MLP-encoded map $f_\theta$ and mean value interpolation $g$. 
Next, we derive the normal vector $\mathbf{n}(\mathbf{x}_j)$ as the cross product of the two directions of the partial derivatives $\partial \mathbf{x}_j / \partial \mathbf{u} = (\partial \mathbf{x}_j / \partial u, \partial \mathbf{x}_j / \partial v)$.

To enforce the smooth connection between two adjacent patches, we employ a \textit{smoothness term}, $L_{\mathrm{smooth}}$, to encourage the normal vectors at this shared smooth connection to align with each other,
\begin{equation}\label{eq:g1_constraint}
    L_{\mathrm{smooth}} = \sum_{\partial \mathcal{S}_{i,j}} \sum_{b} \|\mathbf{n}(\mathbf{x}_{i,b}) - \mathbf{n}(\mathbf{x}_{j,b})\|_2,
\end{equation}
where $\partial \mathcal{S}_{i,j}$ denotes the shared boundary and $b$ indexes a pair of collocated boundary samples within the smooth part of $\partial \mathcal{S}_{i,j}$. As shown in our experiments (c.f. Fig.~\ref{fig:ablation_g1}), this simple practice is sufficient to achieve smooth fitting results at high fidelity.

We define the \emph{boundary fairness} term $L_{\mathrm{fair}}$ based on the curve Laplacian to enforce the fairness along each boundary curve $e$ of the resulting neural parametric patches $\mathcal{S}_i$. We sampled a sequence of points, $\mathbf{x}_{e,j}$, from the boundary curve $e$.
Then we minimize the discrete Laplacians along $\mathbf{x}_{e,j}$,
\begin{equation}
    L_{\mathrm{fair}} = \sum_{e \in \{\partial \mathcal{S}_i \}}\sum_{j} \|\mathbf{x}_{e,j} - (\mathbf{x}_{e,j+1} + \mathbf{x}_{e,j-1})/2\|_2,
\end{equation}
where $\mathbf{x}_{e,j+1}$ and $\mathbf{x}_{e,j-1}$ are the sample points previous and next to $\mathbf{x}_{e,j}$.
This regularization tends to straighten the boundary curves,  so we gradually decrease its influence as the optimization proceeds.

The \emph{uniform parameterization} term $L_{\mathrm{uniform}}$,
introduced in Bedna{\v{r}}{\'\i}k et al.~\shortcite{bednarik2020shape}, is adopted to encourage $E(\mathbf{x}_j) = \partial \mathbf{x}_j/\partial u$ and $G(\mathbf{x}_j) = \partial \mathbf{x}_j / \partial v$ to be similar to their averaged values in the patch. We compute these partial derivatives by differentiating $\mathbf{x}_j$ w.r.t. $\mathbf{u}=(u, v)$ in the 2D parametric domain as before.
The loss term is given as follows:
\begin{equation}
    L_{\mathrm{uniform}} = \sum_{\{\mathbf{x}_j\}} |E(\mathbf{x}_j) - \overline{E}| + |G(\mathbf{x}_j) - \overline{G}|,
\end{equation}
where $\overline{E}$ and $\overline{G}$ are the averaged quantities of $E(\mathbf{x}_j)$ and $G(\mathbf{x}_j)$, respectively.

Finally, we aim to preserve length aspect ratios of the edges of each cell $\mathcal{C}_i$ to make them close to those of $\mathcal{M}_i$. 
For simplicity, we omit the subscript $i$ in the following. We use $|\partial \mathcal{C}^j|$ (or $|\partial \mathcal{M}^j|$) to denote the Euclidean length of the $j$-th boundary curve $\partial \mathcal{C}^j$ of cell $\mathcal{C}$ (or $\partial \mathcal{M}^j$ of $\mathcal{M}$). 
Hence, we define the \emph{aspect ratio loss} as follows:
\begin{equation}
    L_{\mathrm{aspect}} = \sqrt{1 - {d(\mathcal{C})}^T d(\mathcal{M})},
\end{equation}
where $d(\mathcal{C})$ is a normalized vector with its $j$-th entry computed by $d_j(\mathcal{C}) = \frac{|\partial \mathcal{C}^j|}{\sum_{k} |\partial \mathcal{C}^k|}$, and $d(\mathcal{M})$ is likewise defined.

\textbf{Learning from a shape collection.}  
Our proposed pipeline can be applied to model a set of morphable shapes by learning a shape space with our Neural Parametric Surface representation.

To this end, we additionally incorporate a set of learnable latent codes $\{\mathbf{c}_m \in \mathbb{R}^C \}$, each representing a different shape in the collection, and a compact-size MLP network $h_\phi$ to project a given shape latent code $\mathbf{c}_m$ to the feature vectors $\mathcal{Z}_m$ that define the geometry of the feature complex $\mathcal{C}_m$. These added components are shown in Fig.~\ref{fig:pipeline}(b).
We implemented this MLP network $h_{\phi}$ as a broadcast decoder~\cite{watters2019spatial}:  
\[\mathbf{z}^m_k =h_{\phi}(\mathrm{cat}(\mathbf{c}_m, P_k))\]
where $\phi$ are the trainable parameters of $h$, $P_k=1/K$ denotes the positional token for vertex feature $\mathbf{z}_k$, and $\mathrm{cat}(\mathbf{c}_m, P_k)$ denotes the broadcasting concatenation of $\mathbf{c}_m \in \mathbb{R}^C$ and different $P_k$.

As for learning the shape space, our method takes as input a mini-batch of shapes $\mathcal{B}=\{\mathcal{M}_m\}$ and optimizes the latent shape code $\mathbf{c}_m$ for each $\mathcal{M}_m$. We add to Eq.~\ref{eq:reconstruction_loss} a regularization term for the shape latent codes as in~\cite{park2019deepsdf}.  
The total loss for learning a shape space is thus defined as
\begin{equation}\label{eq:learning_space}
    L_{\mathrm{shape}} = \frac{1}{|\mathcal{B}|} \sum_{\mathcal{B}} L_{\mathrm{recon}} + \lambda_{\mathrm{reg}} \sum_{\mathbf{c}_m} \| \mathbf{c}_m \|_2.
\end{equation} 

\section{Experiments and discussions}\label{sec:experiments}

\begin{figure*}[t!]
    \centering
    \begin{overpic}
    [width=0.96\linewidth, trim=0 50 0 0]{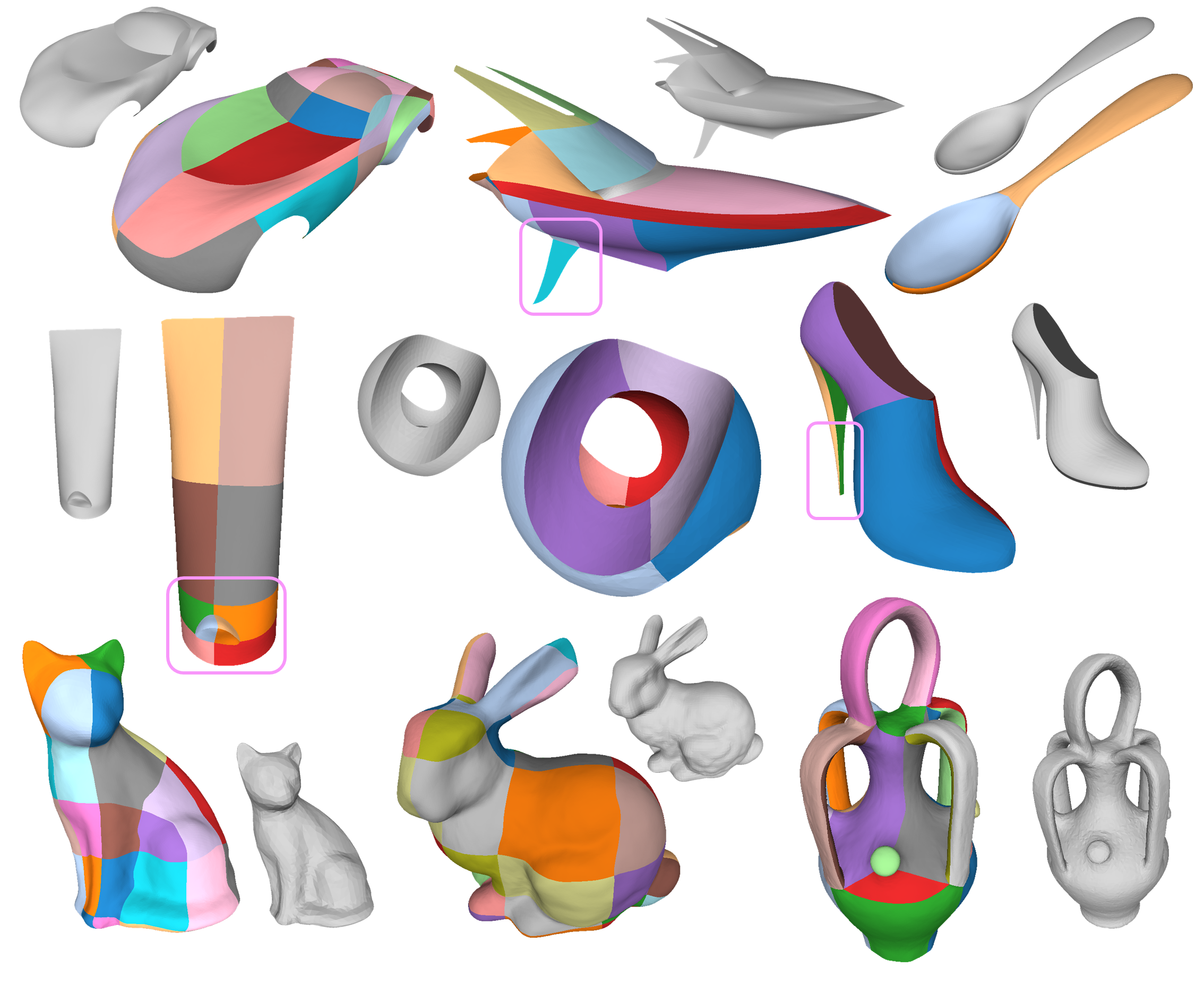}
    \put(2,60){(a) Car}
    \put(35,60){(b) Boat}
    \put(88,60){(c) Spoon}
    \put(2,33){(d) Toothpaste}
    \put(4,31){Tube}
    \put(34,33){(e) Sculpt}
    \put(88,33){(f) High Heel}
    \put(15,0){(g) Cat}
    \put(43,0){(h) Bunny}
    \put(81,0){(i) Vase}
    \end{overpic}
    \caption{
    \textbf{A gallery of surface shapes represented by the proposed neural parametric surfaces}. The target surfaces are shown in grey, and the resulting surfaces are shown with different neural surface patches in different colors. 
    The results show that our method can handle open surface boundaries (in \textit{Car}, \textit{Boat}, and \textit{High heel}), sharp features (in \textit{Boat} and \textit{Sculpt}), and non-manifold patches (in \textit{Boat}). 
    The bottom row shows that our representation can faithfully represent organic shapes, and can work with patch layouts produced by existing tools to produce high-quality results.
    }
    \label{fig:gallery}
\end{figure*}
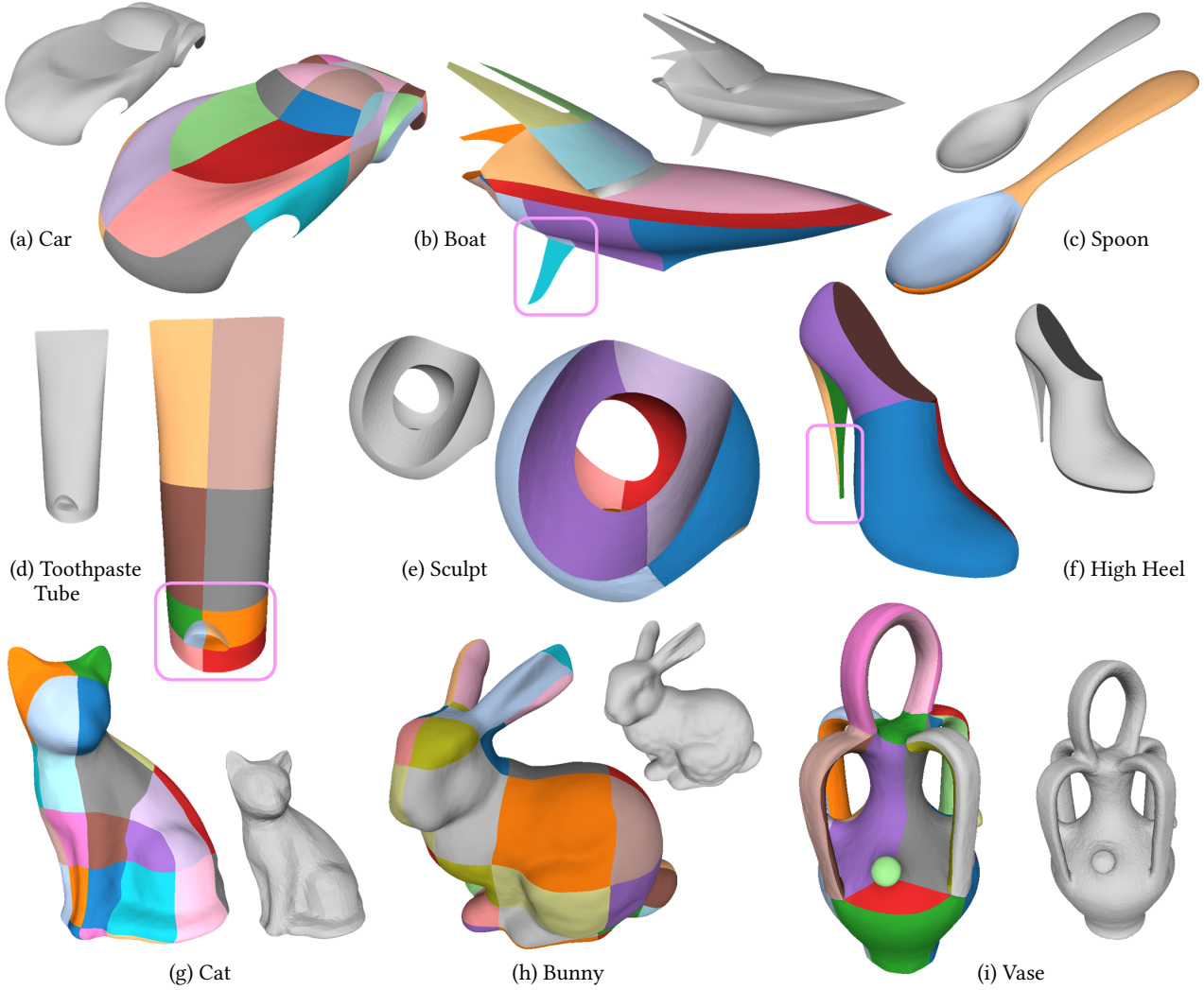

\begin{table*}[h]
    \centering
    \caption{\textbf{The statistics of each shape presented in the paper (the upper block) and the quantitative performance of our method on each shape (the bottom block).} 
    $|\mathcal{Z}|$ and $|\mathcal{S}|$ denote the number of vertices and the number of patches in the given patch partitioning of the shape. 
    $|n \geq 5|$ refers to the number of patches having 5 or more sides and $n_{max}$ refers to the maximal sides of a patch in the given partitioning. \textit{Partitioning} indicates how the patch layout of each shape is generated, with $O$ referring to \textit{layout from original data benchmark}, $M$ \textit{manually prepared}, and $A$ \textit{automatically generated}. For all performance metrics, smaller values are better.
    \vspace{-8pt}    
    }
    \label{tab:statistics}    
    \scalebox{0.80}{
    \begin{tabular}{c| c c c c c c c c c c c c c c c c c c }
    \hline    
         &  Boat & Car & TP-Tube & High heel & Bunny & Cat & Fandisk & Sculpt & Kettle & Catcher & Shoe & Skull & Vase & Spoon & Pants & Skirt & Jeans \\
        \hline
        Partitioning & O & O  & O & O & A & A & A & A+M & M & M & M & M & A+M & M & M & M & M\\
        $|\mathcal{S}|$ & 33 & 36  & 24 & 11 & 59 & 46 & 14 & 10 & 16 & 10 & 13 & 15 & 28 & 6 & 4 & 4 & 4\\
        $|\mathcal{Z}|$ & 44 & 43  & 24 & 16 & 58 & 45 & 24 & 24 & 27 & 20 & 30 & 26 & 45 & 5 & 13 & 9 & 12\\
        $n_{max}$ & 8 & 5  & 5 & 5 & 7 & 5 & 9 & 9 & 9 & 10 & 18 & 13 & 11 & 3 & 6 & 5 & 6 \\
        $|n \geq 5|$ & 10 & 10  & 4 & 4 & 12 & 4 & 6 & 10 & 8 & 3 & 10 & 6 & 17 & 0 & 4 & 2 & 4\\
        \hline
        P2S ($1\times10^{-3}$) & 0.261 & 0.299 & 0.266 & 0.762 & 1.347 & 0.619 & 0.572 & 1.255 & 0.462 & 0.267 & 0.983 & 0.570 & 1.130 & 0.444 & 1.351 & 4.705 & 3.207 \\
        HD ($1\times10^{-3}$) & 3.502 & 5.910 & 2.074 & 11.995 & 15.971	& 6.450 & 8.743 & 8.707 & 4.743 & 3.840 & 14.520 & 6.502 & 18.316 & 3.977 & 18.269 & 89.859 & 36.936 \\
        NAE (deg) & 1.34 & 1.50 & 1.34 & 2.99 & 5.39 & 4.18 & 1.79 & 2.53 & 2.22 & 1.87 & 3.37 & 2.91 & 5.96 & 3.33 & 7.80 & 22.46 & 22.65 \\
        \hline
    \end{tabular}
    }
\end{table*}

\begin{figure*}
    \centering
    \begin{overpic}
    [width=0.96\linewidth, trim = 0 400 0 0]{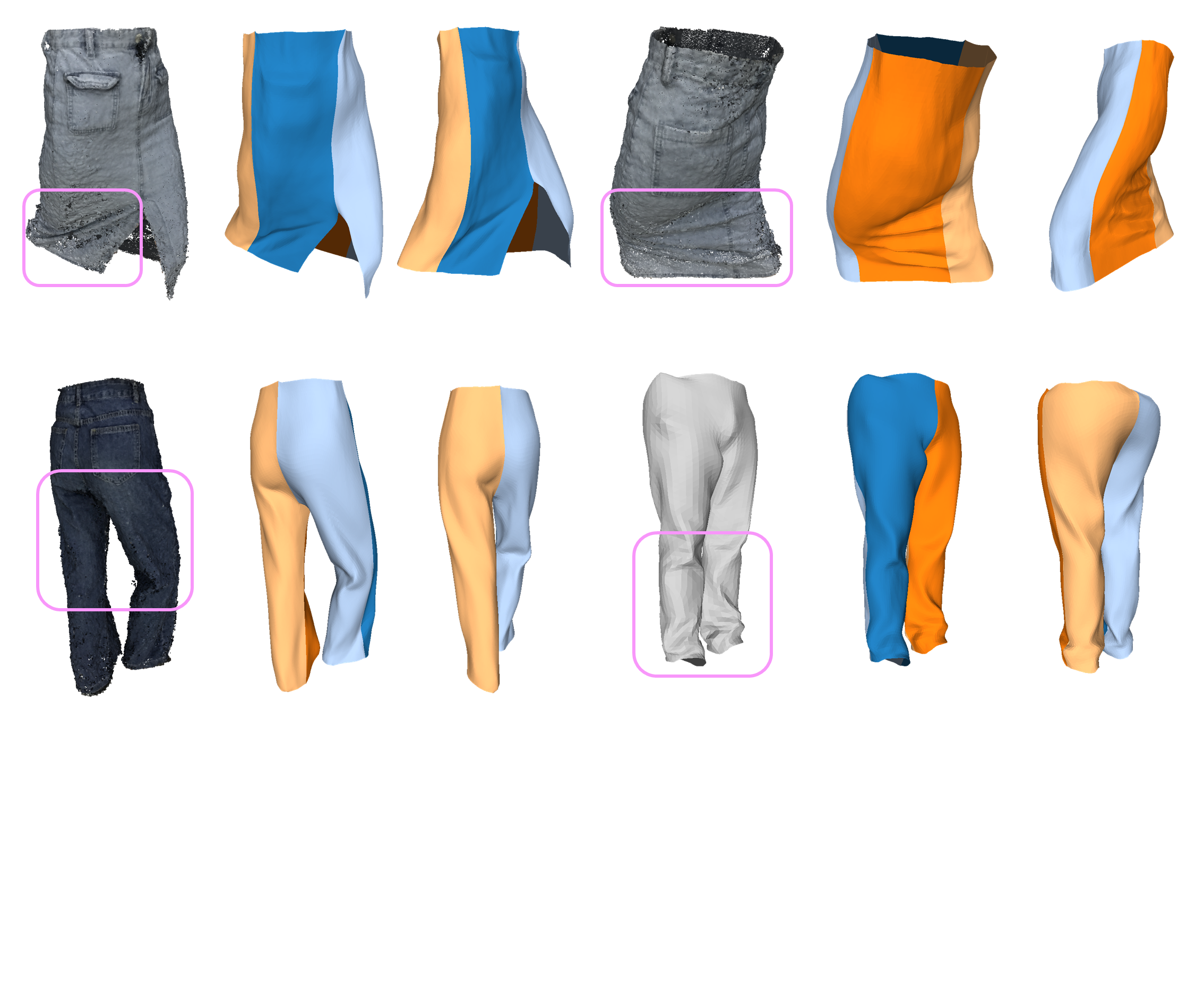}
    \put(27,32){(a) Front panels of the \textit{Skirt} }
    \put(73,32){(b) Back panels of the \textit{Skirt} }
    \put(27,0){(c) \textit{Jeans}}
    \put(73,0){(d) \textit{Pants}}
    \end{overpic}
    \caption{\textbf{Modeling free-form surfaces.} 
    The proposed neural parametric surface representation can fit free-form surface geometries with coarse patch layouts (using just 4 patches for the \textit{Skirt} front panels (a) and back panels (b), \textit{Jeans} (c), and \textit{Pants} (b)) while retaining the geometric details including wrinkles.
    Among them, the \textit{Skirt} and \textit{Jeans} are represented as point clouds.
    In each subfigure, the left is the target geometry to fit; the middle is the reconstructed surface represented by the Neural Parametric Surface from the same view of the target geometry; the right is the fitting result viewed from another viewpoint.
    }
    \label{fig:freeform}
\end{figure*}

\subsection{Implementation details}\label{sec:implementation}

\textbf{Network architecture and training. }
In this work, we use a $128$-dimensional feature complex for each shape. The MLP network implementing the mapping function $f$ has $12$ layers with each hidden layer having $256$ neurons. We employ \textit{softplus} activations ($\beta=100$) in between the layers to ensure differentiability~\cite{gropp2020implicit}. The MLP network for the broadcast decoder is designed to be very compact, with $3$ layers and using the same activation functions. We implement the networks in PyTorch~\cite{NEURIPS2019_9015} and used its \textit{autograd} function to analytically compute partial derivatives $\partial \mathbf{x} / \partial \mathbf{u}$ used in the loss.

For all the experiments on the \textit{shape fitting} task (Sec.~\ref{sec:shapefitting}) shown in this paper, results are obtained with $2,000$ iterations to ensure the optimization convergence for shapes with different levels of complexity. In each training iteration, a batch of $10,000$ points is randomly sampled from the input surface. The number of points sampled from each patch depends proportionally on the area of this patch. In this task, we adopt a warm-up stage to initialize the network with the training objective $L_{\mathrm{recon}} = L_{\mathrm{anchor}} + \lambda_{\mathrm{uniform}} L_{\mathrm{uniform}}$ for the first 100 training iterations. After this warm-up, we minimize the full loss in Eq.~\ref{eq:reconstruction_loss}. After the 300 iterations, we decay the influence of $L_{\mathrm{curvature}}$ to avoid excessive straightening of the boundary curves of the neural parametric patches. The Adam optimizer~\cite{kingma2014adam} is used with an initial learning rate of $10^{-3}$ and a cosine annealing scheduler is employed to gradually decay the learning rate to $10^{-5}$ during the course of optimization. 

All results shown in the experiments concerning the \textit{shape-space learning} task (Sec.~\ref{sec:shapespace}) are obtained with $100$ epochs. In each iteration, a mini-batch of 24 shapes is provided. $5,000$ points are sampled from each shape in a similar practice as described for the shape fitting task. The full loss in Eq.~\ref{eq:reconstruction_loss} is minimized during the entire training course. We use the Adam optimizer~\cite{kingma2014adam} with an initial learning rate of $10^{-3}$ which is decayed to $5\times10^{-4}$ for the final 20 epochs. More implementation details can be found in our open-source code for reproducibility. 

We measure the reported time efficiency of the proposed method on a Linux OS desktop with an NVIDIA GeForce RTX 4090 (24G memory) graphics card and an Intel(R) Core(TM) i7-10870H CPU.

\textbf{Preparation of the patch layout for shapes. }
Many existing methods can be used to generate a patch layout, enabling our approach to fit a \textit{Neural Parametric Surface} to a given surface shape. When the corner vertices of the patch layout are specified by users on the given surface shape, we can use the method in~\cite{born2021layout} to automatically generate a quad layout, which is compatible with our method. For generating a patch layout that allows $n$-sided polygonal patches, we can use tools such as LoopyCuts~\cite{livesu2020loopycuts} or Variational Shape Approximation (VSA)~\cite{cohen2004variational}.

The \textit{only} requirement for our patch layout is that each patch should be a 2-manifold with a \textit{disc} topology, and each arc should be uniquely defined by its vertices. Namely, the graph constructed from the corner points and arcs of the patch layout should be a \textit{simple} undirected graph. In this graph, each arc should connect two distinct corner points, thereby avoiding \textit{self-loops}; and there should be a maximum of one arc between any pair of corner points, thereby precluding \textit{multiple edges}. This requirement ensures correct linear interpolation in the feature space. A semantic segmentation result may contain self-loops or multiple edges, and violate this requirement, hence, cannot be directly used here. To address this, we have developed an interactive segmentation tool to cut \textit{self-loops} into valid arcs or insert new corner points to remove \textit{multiple edges}.

In our experiments, we show our method can robustly work with patch layouts obtained from different methods. For example, models such as \textit{Car}, \textit{Boat}, and \textit{Toothpaste tube} use layouts from an existing dataset~\cite{pan2015flow,bae2008ilovesketch}. Models like \textit{Pants}, \textit{Skirt}, and \textit{Jeans} use layouts manually prepared by garment designers similar to~\cite{pietroni2022computational}. For the \textit{Bunny} and \textit{Cat} models, their patch layouts are automatically generated using LoopyCuts ~\cite{livesu2020loopycuts}. Subsequent results will demonstrate that our method is not sensitive to different patch layouts, whether they are prepared by LoopyCuts or VSA~\cite{cohen2004variational}.

\textbf{Meshing neural parametric patches for visualization. }
Since our representation is inherently a \textit{parametric} one, visualizing the surface geometry requires no more than tessellating the neural parametric surface patches. To this end, we sampled both inside the parametric domain $\Omega_i$ and along its boundary $\partial \Omega_i$ to obtain a set of planar coordinates $\{\mathbf{u}_j\}$. We tessellated these 2D samples with a constrained Delaunay triangulation. This triangulation in the parametric domain is then used to tessellate $\mathbf{x}_j$ in the resulting neural parametric patch in 3D space. A subsequent local curvature-based edge-flipping algorithm is applied to avoid aliasing. The runtime cost of the entire visualization pipeline depends on the density of the output mesh. For example, the pipeline runs at approximately 10 frames-per-second (FPS) when generating a mesh surface with $12,000$ vertices. Note that this visualization pipeline is not different from traditional pipelines for rendering parametric surfaces.
This stands in contrast to pipelines designed for rendering implicit representation based on ray casting or Marching Cubes algorithms~\cite{lorensen1998marching} that contour the zero-level set.

\subsection{Surface fitting with neural parametric surfaces}\label{sec:shapefitting}

\subsubsection{Data preparation and metrics}
We evaluated our proposed Neural Parametric Patches on various shapes~\cite{bae2008ilovesketch,pan2015flow,zhu2020deep,bhatnagar2019multi}. All shapes are normalized such that they are centered at the origin and the maximal extent of the shape is 2. Table~\ref{tab:statistics} summarizes the statistics of each shape, including the number of patches $|\mathcal{S}|$, the number of vertices $|\mathcal{Z}|$ of the feature complex, the number of $n$-sided patches where $n \ge 5$, and the maximal $n$ in a patch layout.

\emph{Fitting accuracy} were evaluated using three metrics: 
\begin{enumerate}
    \item A two-sided point-to-surface (P2S) distance that measures the average distance between the reconstruction and the target; 
    \item The two-sided Hausdorff distance (HD) that measures the largest discrepancy between the reconstructed and target shapes; and
    \item Averaged angular (degree) error (NAE) between normal vectors of corresponding points (i.e.\ the closest points, as described in Sec.~\ref{sec:training}).
\end{enumerate}
All reported metrics are computed with $30,000$ points randomly from both the target and the reconstructed surfaces.

\subsubsection{Experimental results}

Fig.~\ref{fig:gallery} shows a gallery of the shapes represented by the proposed Neural Parametric Surface. Our Neural Parametric Surface can convert a surface shape to a seamless parametric surface representation given a layout partitioning the shape. The neural parametric surfaces are shown in colors with each color indicating a surface patch and the target surfaces in grey are provided next to the results.

For example, the \textit{Car} model shows the proposed representation retains the \textbf{smoothness} of the car surface of the target surface in grey. In addition, our representation can model \textbf{non-manifold surface patches} and \textbf{open surfaces} well (e.g.\ the fin keel, that is the turquoise patch at the bottom of the \textit{Boat} model). It can reproduce \textbf{sharp creases} (e.g.\ the enclosed region of the \textit{Toothpaste Tube} model and the sharp edges in the \textit{Sculpt} model). It is also able to represent \textbf{thin geometry}, such as the thin slender \textit{Spoon} model or the slender heel of the \textit{High heel} model.

Such geometric features are commonly seen in applications related to CAD/CAM, reverse engineering, or product design. 
Although traditional parametric representations can handle such cases, they require a high level of design expertise to ensure seamless results, due to the topological constraints of the parametric domain. Such features are also very challenging to model with neural implicit representations. Our method, however, naturally supports seamless handling of such features, extending the domain of neural representations, and easing the modeling burden compared to traditional parametric approaches. 

More results in the bottom row of Fig.~\ref{fig:gallery} show that our representation can approximate organic objects at high fidelity. Among them, the patch layouts of the \textit{Cat} and the \textit{Bunny} are prepared by LoopyCuts. From the results, we see that our proposed representation can work along with existing layout generation methods to produce high-quality results.

\textbf{Modeling shapes with coarse patch layouts.} 
Our proposed neural parametric surface representation can also model \emph{free-form shapes with very coarse patch layouts}. Both the \textit{Skirt} and \textit{Jeans} of Fig.~\ref{fig:freeform}, and the \textit{Pants} of Fig.~\ref{fig:teaser} are modeled with just four surface patches. Despite the coarse patch layout, geometric details such as the wrinkles at the bottom of both shapes are well captured with a single neural parametric patch. Further, note how wrinkles are smoothly reconstructed across adjacent patch boundaries.  This demonstrates the advantage of our proposed representation brought by the deep neural networks, avoiding local refinement required by traditional parametric surface representations that result in dense control meshes.

\textbf{Reconstructing incomplete shapes.} 
We also note that our approach can utilize the given topological partitioning of the target geometry to approximate the target geometry with strong resilience to \emph{imperfect data}. For example, in Fig.~\ref{fig:freeform}, both the \textit{Skirt} and \textit{Jeans} models are point clouds from multiple scans~\cite{zhu2020deep}, which suffer from missing data. Our result can produce plausible outcomes by smoothly filling in the region where ground-truth data is missing. 

\textbf{Fitting accuracy and runtime statistics.}
Tab.~\ref{tab:statistics} reports the fitting errors in terms of the three metrics introduced earlier. Our method achieved very high fitting accuracy as compared to existing neural patch-based approaches; see Tab.~\ref{tab:comparison}. The \textbf{training time} needed for the listed shapes ranges from $\sim$10 minutes (e.g.\ for the \textit{Pants}) to around half an hour (e.g.\ for the \textit{Bunny} with the most number of patches), that is comparable to similar methods. The \textbf{storage} that our representation takes is around $5.5$ MB for all listed shapes, whereas a meshed model containing $200k$ vertices requires around $20$ MB for storage. Compared with other neural approaches (c.f.\ Tab.~\ref{tab:comparison}), our representation is more compact. The limited use of storage is owing to the fact that the increase of the patch complexity incurs a little additional storage requirement for the feature complex, while the parameters used to represent mapping function $f$ stay unchanged. This shows that our method is scalable to very complicated shapes and thus can be used as a compression tool for surface geometry. 

In summary, these experiments demonstrated that the Neural Parametric Surface is a flexible representation that can handle complicated geometry with a variety of geometric features. It can model a given shape using coarse patch layouts with high fidelity and is compact in terms of storage.

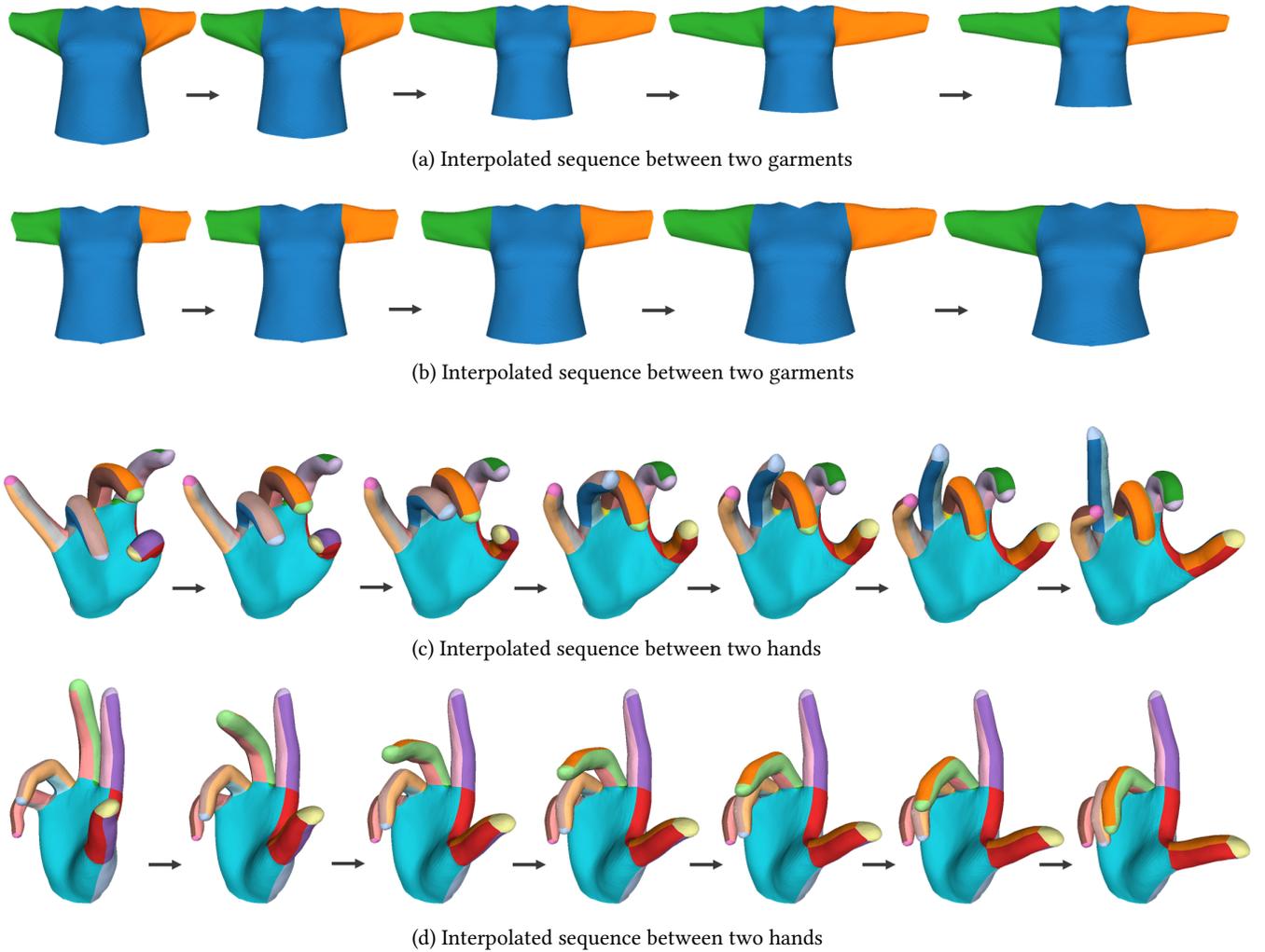
\begin{figure*}
    \centering
    \begin{overpic}
    [width=\linewidth, trim = 0 100 0 0, clip]{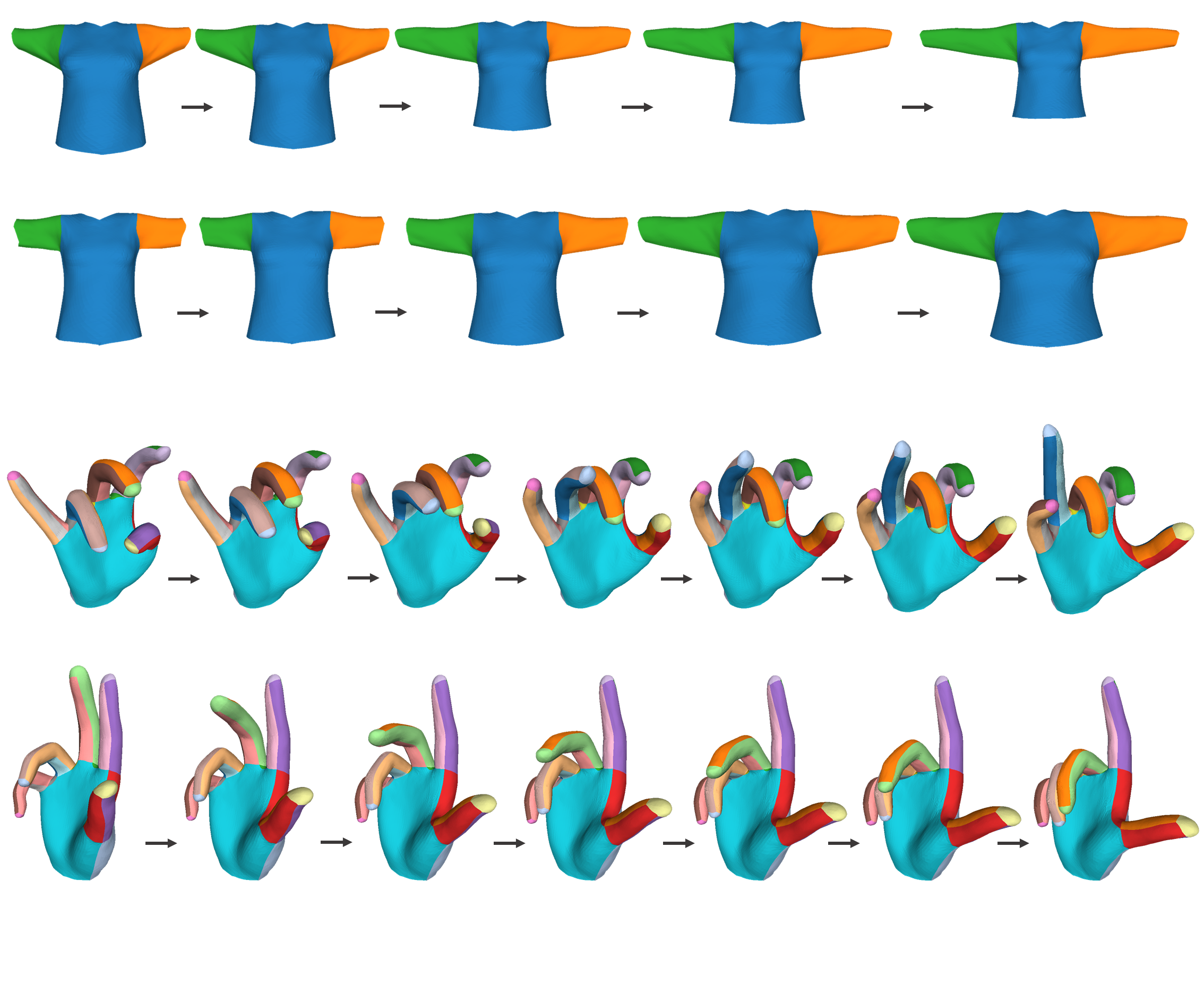}
    \put(33,63){(a) Interpolated sequence between two garments}
    \put(33,46){(b) Interpolated sequence between two garments}
    \put(33,24){(c) Interpolated sequence between two hands}
    \put(33,1){(d) Interpolated sequence between two hands}
    \end{overpic}
    \caption{\textbf{Interpolated sequences between randomly selected pairs of shapes.} 
    (a) and (b) show interpolated sequences between two pairs of garments (shown at the left and right ends in each individual row), which have different styles and sizes. The garments are aligned to their collar to better visualize the size change.
    (c) and (d) show the interpolated sequences between two pairs of hands with different pose codes. 
    Please refer to the supplementary video for the entire sequence of the interpolated results.}
    \label{fig:interpolation}
\end{figure*}

\subsection{Neural Parametric Surface for Sets of Shapes}\label{sec:shapespace}

Our proposed Neural Parametric Surface representation can learn a morphable shape space from a set of 3D shapes. In the following, we show several applications that demonstrate the capability and potential of this proposed representation. 

\textbf{Datasets. } 
We utilize two datasets for our experiments: the first consists of various types of T-shirts from a garment dataset~\cite{wang2018learning}, and the second includes $10,000$ randomly sampled hands from a large-scale hand dataset~\cite{gao2022dart}. For the \textit{garment} dataset, we align all 3D meshes through their collars by translating the meshes so that the centers of collars coincide with the origin. Then, we scale these meshes by a constant factor to normalize the meshes. Similarly, for the \textit{hand} dataset, we align all meshes to their roots and scale them with a constant factor to normalize their size.

\subsubsection{Shape interpolation in learned latent space.}
We show a shape interpolation application using the latent space learned from the garment dataset. We employ an auto-decoding approach similar to that in \cite{park2019deepsdf}. Specifically, each garment is represented by a unique, learnable latent code $\mathbf{c}_m$. A broadcast network $h_{\phi}$ is employed to map this latent code $\mathbf{c}_m$ to a feature complex $\mathcal{C}_m$. After this broadcast decoding step, we feed samples from the feature complex $\mathcal{C}_m$ to the MLP-encoded map $f_{\theta}$ to produce a 3D reconstruction that fits the garment corresponding to $\mathbf{c}_m$. This entire pipeline (shown in Fig.~\ref{fig:pipeline}) is trained end-to-end by minimizing the loss defined in Eq.~\ref{eq:learning_space}. 

After training the networks $h$ and $f$, we obtain a codebook containing latent codes that represent various garment shapes. To perform interpolation between two specific garment shapes, we first randomly select two latent codes from the codebook, then generate a sequence of interpolated latent codes using linear interpolation between the two selected codes. Next, these interpolated latent codes are fed into the pipeline comprising $h$ and $f$ to generate a series of 3D garment geometries. Two sequences of interpolated garments are shown in Fig.~\ref{fig:interpolation}(a) and (b). The interpolation smoothly morphs the two end shapes without introducing hollowing or undesirable artifacts (see the supplementary video for detailed results). This shows that the proposed method can learn a meaningful morphable shape space for design exploration. 

\begin{figure}
    \centering
    \begin{overpic}
    [width=\linewidth, trim=0 500 0 0, clip]{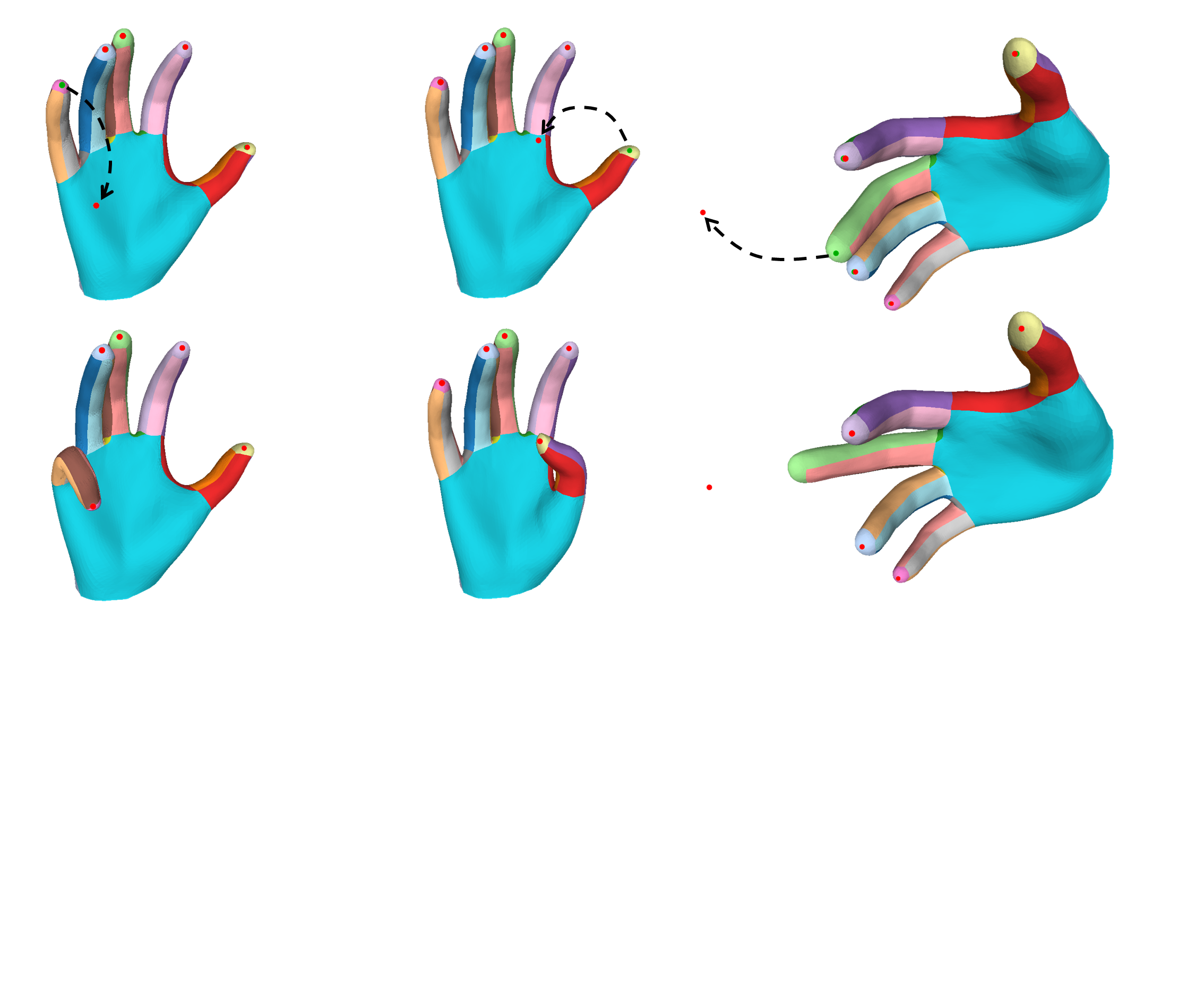}
    \put(8,0){(a)}
    \put(40,0){(b)}
    \put(80,0){(c)}
    \end{overpic}
    \caption{\textbf{Editing hand meshes with sparse handles.} 
    The top row shows the rest pose of each example, and the bottom row shows the deformation results. The red dots represent the target positions of the fingertips. The root of the hand is also fixed during the test-time optimization. Example (c) shows an extreme case where the target position of the middle fingertip is set to be far away from the hand.}
    \label{fig:hand_editing}
\end{figure}

\subsubsection{Editing hand meshes with sparse handles.}
We showcase another application of our proposed representation by deforming a hand mesh. Specifically, based on a learned latent space of hand shapes, we allow interactive editing on one or more fingertips. The training pipeline is identical to the one described previously, and is trained by minimizing the loss function given in Eq.~\ref{eq:learning_space}, but applied to a dataset of hand shapes.

At the runtime stage, we allow users to specify the target positions of the fingertips. Each fingertip is represented as the center of the fingertip region. To deform the hand mesh, we use the source and target positions of the fingertips as deformation handles, and solve the deformation problem as a test-time optimization problem:
\begin{equation}\label{eq:hand_editing}
    L = \argmin_{\mathbf{c}_m} \|\mathrm{cog}(\{\mathbf{x}\}_i) - \mathbf{p}_i\| + 10^{-3}\|\mathbf{c}_m\|_2,
\end{equation}
where $\{\mathbf{x}\}_i$ denotes the sample points from the neural parametric patch corresponding to the $i$-th fingertip and $\mathrm{cog}(\{\mathbf{x}\}_i)$ denotes the center of this region; $\mathbf{p}_i$ is the user-specified target position; and $\mathbf{c}_m$ denotes the trainable latent code that defines the hand pose. The only variable is $\mathbf{c}_m$, while all the networks $h$ and $f$ are fixed.

We adopt the ADAM optimizer (with default parameters) and a constant learning rate of $0.005$ to solve the above problem (Eq.~\ref{eq:hand_editing}) at an interactive rate (15 Hz). Adding the regularization term (the second term in Eq.~\ref{eq:hand_editing}) can effectively enforce an intuitive deformation. We show three exemplary results of hand meshes that are edited by posing their fingertips in Fig.~\ref{fig:hand_editing}. In the first two examples Fig.~\ref{fig:hand_editing}(a) and (b), the little finger and the thumb are specified to bend towards the palm, and the deformation results of the hand shapes are intuitive. We also demonstrate the robustness of this approach by illustrating how it handles the bending and over-stretching of a finger, as shown in Fig.~\ref{fig:hand_editing}(c). Even when the user-specified target position extends beyond the finger's natural range of motion, our pipeline still produces a plausible deformation, guiding this (middle) finger toward the target position without introducing undesirable artifacts.

\begin{figure*}
    \centering
    \begin{overpic}
        [width=\linewidth, trim = 0 700 0 0]{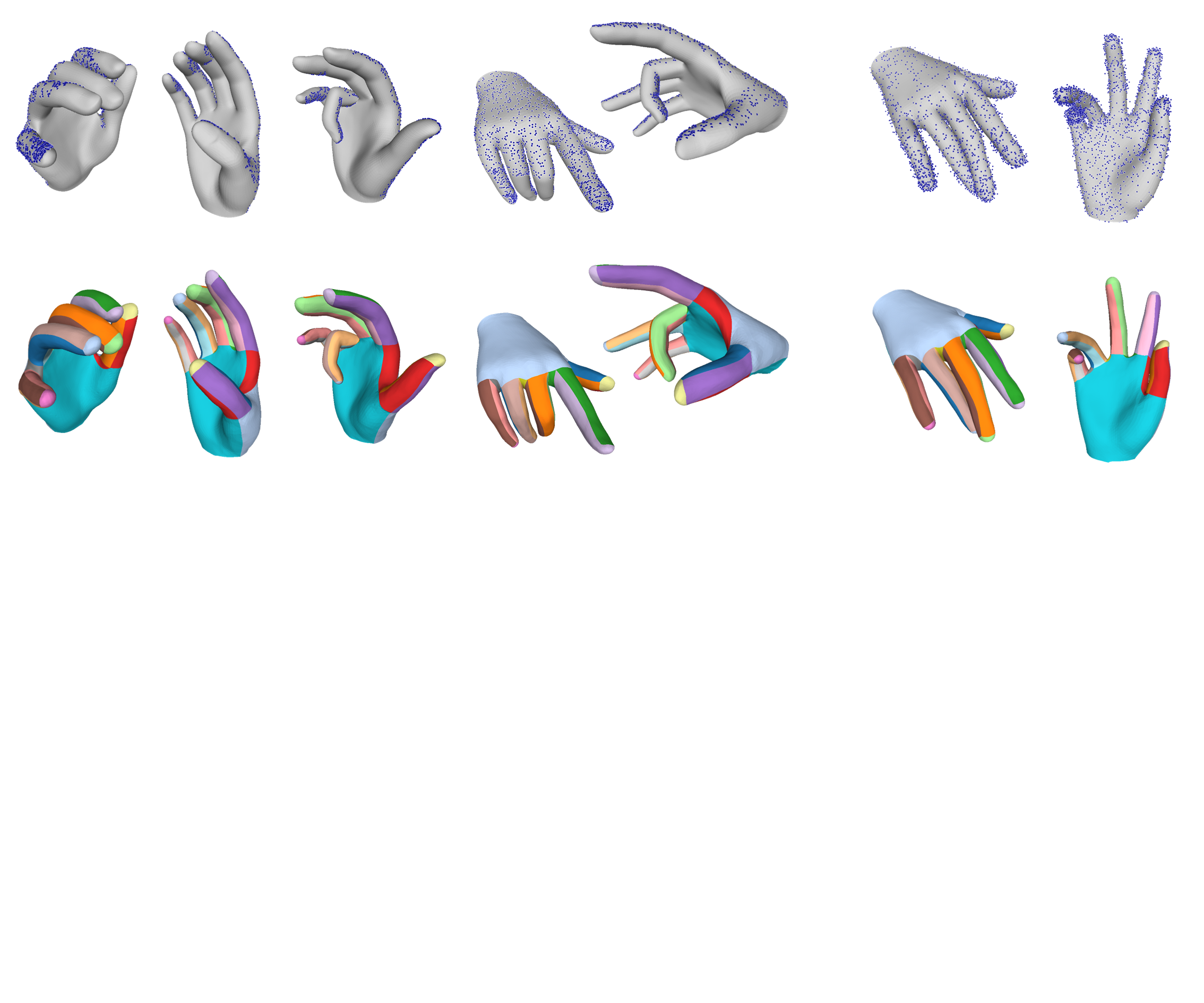}
        \put(7,0){(a) Partial point cloud (blue points) simulating single view scans.}
        \put(73,0){(b) Noisy point cloud (blue points).}
    \end{overpic}
    \caption{\textbf{The neural parametric surface augmented with a learned latent shape applied to reconstruct point cloud data without layout information}. 
    In the 1st row, blue points are the input partial (a) or noisy (b) point clouds; the ground-truth (GT) models are provided in grey for reference. 
    (Note: GT hands in grey are not used for reconstruction; they are only shown as a reference for better visualization.)  
    The 2nd row shows the fitting results obtained by our method, with different colors indicating different surface patches. In most cases, our model reconstructs the ground-truth poses with high quality. 
    }
    \label{fig:registration}
\end{figure*}

\subsubsection{Fitting imperfect point cloud data.}\label{sec:template_fitting}
Once trained on a large set of shapes, our Neural Parametric Surface can be used to generate a surface from imperfect point cloud data without needing the patch layout information. We showcase this application on the hand dataset. To prepare the imperfect point cloud data for a given hand mesh, we introduce two types of imperfections: one arising from noisy scans and the other from partial observations due to a single-view scan. To simulate the noisy point cloud, we perturb the surface sample points and their normals on a ground-truth (GT) hand mesh by adding Gaussian noise $\mathcal{N}\sim(0, 0.01^2)$. To replicate a single-view scan, we position a viewpoint one unit away from the origin, facing the center of the hand's palm, and then employ a simple strategy to filter out sample points whose normal directions form acute angles with the viewing direction. 

We train a PointNet-based predictor~\cite{qi2017pointnet++} on the hand dataset to predict the latent shape code $\mathbf{c}_m$ from a given imperfect point cloud data. During runtime, we first obtain an initial guess of the shape code $\mathbf{c}$ of a given point cloud data. Then, we perform a test-time optimization to optimize the shape code $\mathbf{c}$ so the generated neural parametric surface fits the given point cloud. The optimization problem is similar to Eq.~\ref{eq:hand_editing} with slight modification as follows:
\begin{equation}
    L = \argmin_{\mathbf{c}} \sum_{(\mathbf{x}_i, \mathbf{p}_i)}L_{\mathrm{surface}} + L_{\mathrm{normal}} + 10^{-3}\|\mathbf{c}\|_2,
\end{equation}
where the pairs of $(\mathbf{x}_i, \mathbf{p}_i)$ are corresponding point pairs derived as described for Eq.~\ref{eq:surface_fitting} with an additional normal filtering scheme to filter out mismatches. Specifically, we filter out the point pairs whose cosine similarity values between the normal vectors at the respective paired points are larger than $0.7$. The fitting results are shown in Fig.~\ref{fig:registration}. Our proposed representation can robustly handle imperfect data from single-view scans (Fig.~\ref{fig:registration}(a)) and noisy inputs (Fig.~\ref{fig:registration}(b)). This validates our pipeline based on the proposed Neural Parametric Surfaces can learn a plausible shape space useful for the deformable template fitting task.

\subsubsection{Mapping hand poses to hand meshes.}
Lastly, we demonstrate that our Neural Parametric Surface representation can function as a \textit{mapping mechanism} to translate from abstract, physically meaningful concepts such as hand poses, to complex surface geometries like hand meshes. We adopt the same training strategy as used in the previous applications; the \textit{only} difference is to replace the learnable latent codes with the pose vectors of the hand meshes in the dataset. To showcase the capability of our approach, we randomly select a pair of hand pose vectors from the dataset and calculate interpolated pose vectors through linear interpolation.\footnote{Since the pose parameters are represented as the rotational angles, we rule out the sequences that contain invalid poses from the interpolation.} Our pipeline takes as input a valid pose vector and generates a hand mesh in the Neural Parametric Surface representation through the broadcasting network $h$ and then the geometric decoder $f$. Examples of the generated hand meshes are shown in Fig.~\ref{fig:interpolation}(c) and (d). These results demonstrate the proposed pipeline can serve as a pose-to-shape mapping function for explicit shape editing.

\begin{figure*}
    \centering
    \begin{overpic}
    [width=\linewidth, trim=0 430 200 0, clip]{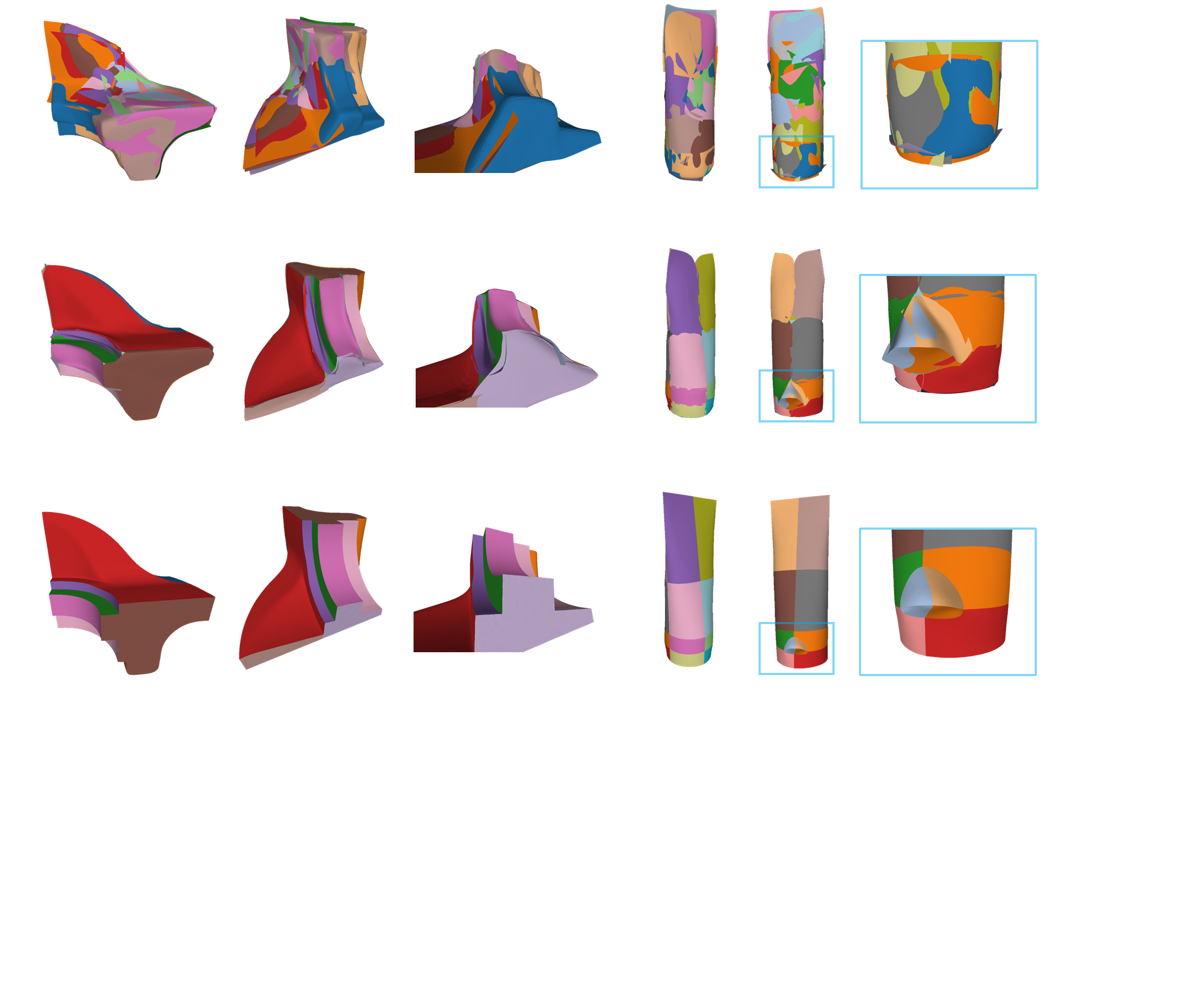}
    \put(45,0){(c) Our results.}
    \put(20,23){(b) Results produced by DSP-BetterStitch$^*$ \textit{with} surface partitioning supervision.}
    \put(20,45){(a) Results produced by DSP-BetterStitch \textit{without} surface partitioning supervision.}
    \end{overpic}
    \caption{\textbf{Qualitative comparison among Deng et al.~\shortcite{deng2020better} 
    (DSP-BetterStitch), its modified version (DSP-BetterStitch$^*$), and our results.}
    DSP-BetterStitch$^*$ (middle row) failed to generate connected surface geometry even for the rectangular patches (upper part of the \textit{Toothpaste Tube}). The use of a rectangular domain also makes it difficult to handle $n$-sided patches (see the zoom-in view). 
    Our neural complex map can faithfully reproduce the seamless surface geometry with the surface partitioning preserved.      
    }
    \label{fig:comparison}
\end{figure*}

\subsection{Comparison with existing works}

\subsubsection{Comparison with neural patch-based methods}
Most existing patch-based neural representations build upon a parametric atlas and model a given surface with a set of unorganized, overlapped patches to reconstruct the geometry. Therefore, they are not directly comparable to our representation that aims at not only reproducing the geometry with $G^0$-continuity but also maintaining the structural information inherent to the given semantics. We compare our patch-based neural representation with Deng et al.~\shortcite{deng2020better}, which is most relevant to this work. Building on AtlasNet~\cite{groueix2018papier}, Deng et al.~\shortcite{deng2020better} improves the differential surface representation (DSP)~\cite{bednarik2020shape} with a stitching loss term to minimize the gap between spatially close patches. We denote this method as \textit{DSP-BetterStitch} in the following.

To establish a fair comparison, we provide to \textit{DSP-BetterStitch} both the 3D geometry and the corresponding partitioning and require each of its parametric patches to fit one of the partitioned patches. We denote this modified version as \textit{DSP-BetterStitch$^*$}. 

\textbf{Comparison results.}  Fig.~\ref{fig:comparison} compares our method with \textit{DSP-BetterStitch} and \textit{DSP-BetterStitch$^*$} (the adapted version) on two shapes: the \textit{Fandisk} and \textit{Toothpaste Tube}. Both shapes have multiple rectangular patches in the given partitioning (8 out of 14 for \textit{Fandisk} and 12 out of 24 for \textit{Toothpaste Tube}). \textit{Fandisk} also has several 9-sided patches.

\begin{table}[]
    \centering
    \caption{\textbf{Fitting error comparison between our and other methods.} DSP-Stitch refers to~\cite{deng2020better}, DGP refers to~\cite{williams2019deep}, DSP-Stitch$^*$ refers to results derived with~\cite{deng2020better} supervised by the given patch decomposition (for a fair comparison with our method).
    }
    \scalebox{0.8}{
    \begin{tabular}{c c|c c c c }
    \hline
         &  & P2S & HD & NAE & NN \\
         Method & Shape &  $\times 10^{-3}$ & $\times 10^{-3}$ & (deg) & Storage \\
        \hline
        \multirow{2}{*}{DSP-BetterStitch$^*$}
            & Fandisk & 5.44 & 58.53 & 14.32 &  22.3 MB \\
            & T-Tube & 3.39 & 55.00 & 7.39 &  38.2 MB \\
        \hline
        \multirow{2}{*}{DSP-BetterStitch}
            & Fandisk & 5.34 & 67.99 & 14.29 & 22.3 MB \\
            & T-Tube & 1.73 & 29.13 & 5.90 & 38.2 MB \\
        \hline
        \multirow{2}{*}{DGP}
            & Fandisk & 1.13 & 56.17 & 7.56 & 115 MB (67 patches) \\
            & T-Tube & 0.41 & 20.19 & 3.53 & 80.8 MB (47 patches)\\
        \hline
        \multirow{2}{*}{Ours}
            & Fandisk & \textbf{0.57} & \textbf{8.74} & \textbf{1.79} & 5.5 MB \\
            & T-Tube & \textbf{0.26} & \textbf{2.07} & \textbf{1.34} & 5.5 MB \\
        \hline
    \end{tabular}}
    \label{tab:comparison}
\end{table}

\textit{DSP-BetterStitch$^*$} can produce semantically aligned results given the surface partitioning as supervision. However, the seams between neighboring patches are not well organized to stitch the patches. Instead, they often intersect with each other (as shown in \textit{Fandisk}) or severely overlap (as shown in \textit{Toothpaste Tube}). These unpleasing results can be observed even for the patches with rectangular partitioning (see \textit{Toothpaste Tube}), and could lead to undesirable holes around the intersection of multiple patches in the target shape. 

We also provide for reference purposes the results by the original implementation of \textit{DSP-BetterStitch} using the same number of patches (12 for \textit{Fandisk} and 24 for \textit{Toothpaste Tube}). While the collection of parametric patches can capture the rough geometry of the given shapes, it is difficult for this method to reproduce meaningful features presented in the target shapes (see the zoom-in view for \textit{Toothpaste Tube} where the feature with sharp creases is missing).

Unlike \textit{DSP-BetterStitch} and \textit{DSP-BetterStitch$^*$}, our method can reconstruct the given shape following its original patch partitioning. Furthermore, it can nicely preserve geometry smoothness and sharp crease lines.

\begin{figure}
    \centering
    \begin{overpic}
    [width=\linewidth, trim = 0 850 850 0, clip]{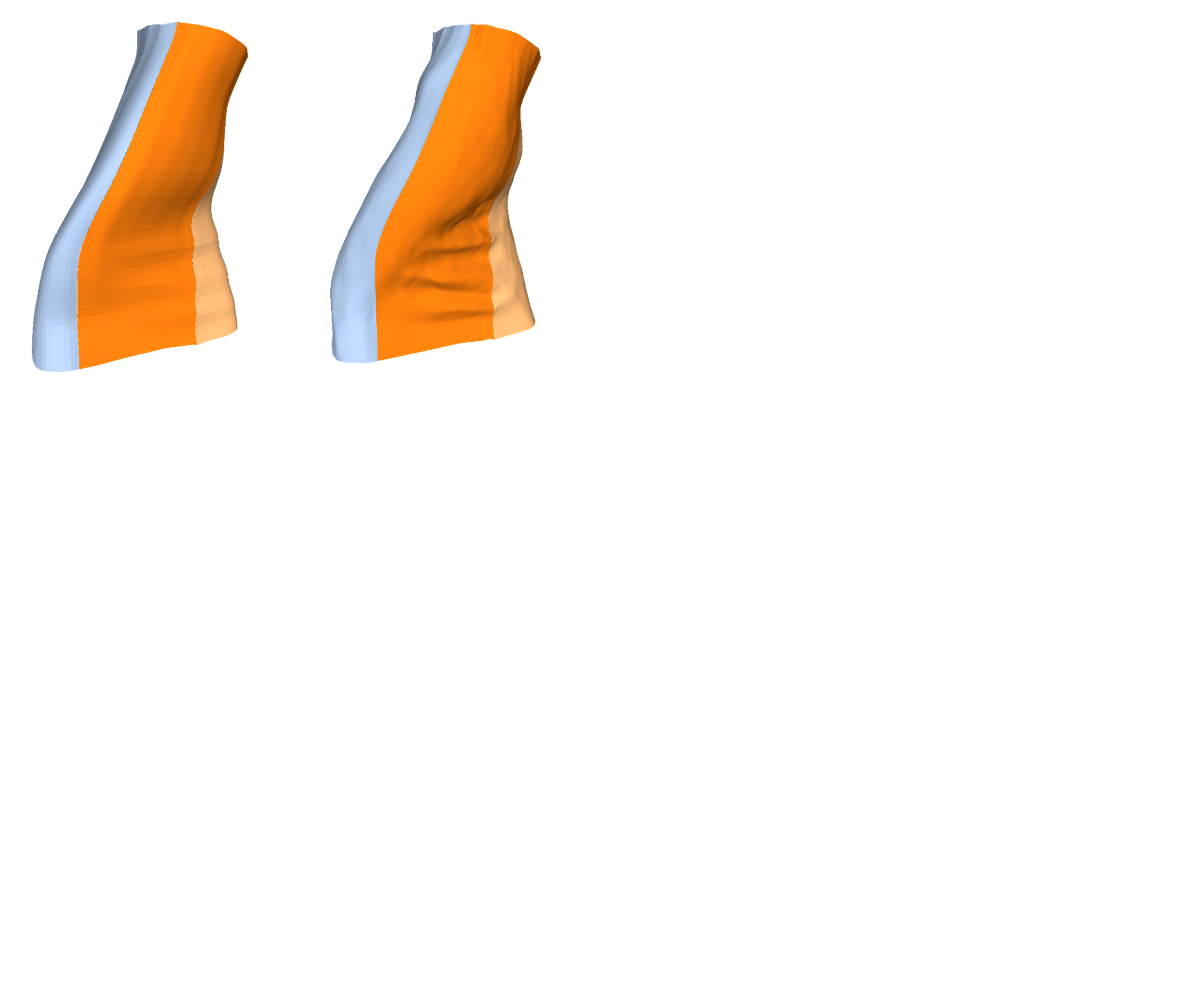}
    \put(7,0){(a) Coons Patches}
    \put(47,0){(b) Neural Parametric Surface}
    \end{overpic}
    \caption{Comparison between the Skirt models produced by Coons patches (a) and by our method (b). }
    \label{fig:comp_w_coons}
\end{figure}

\begin{figure}
    \centering
    \begin{overpic}
    [width=\linewidth, trim = 0 0 0 0, clip]{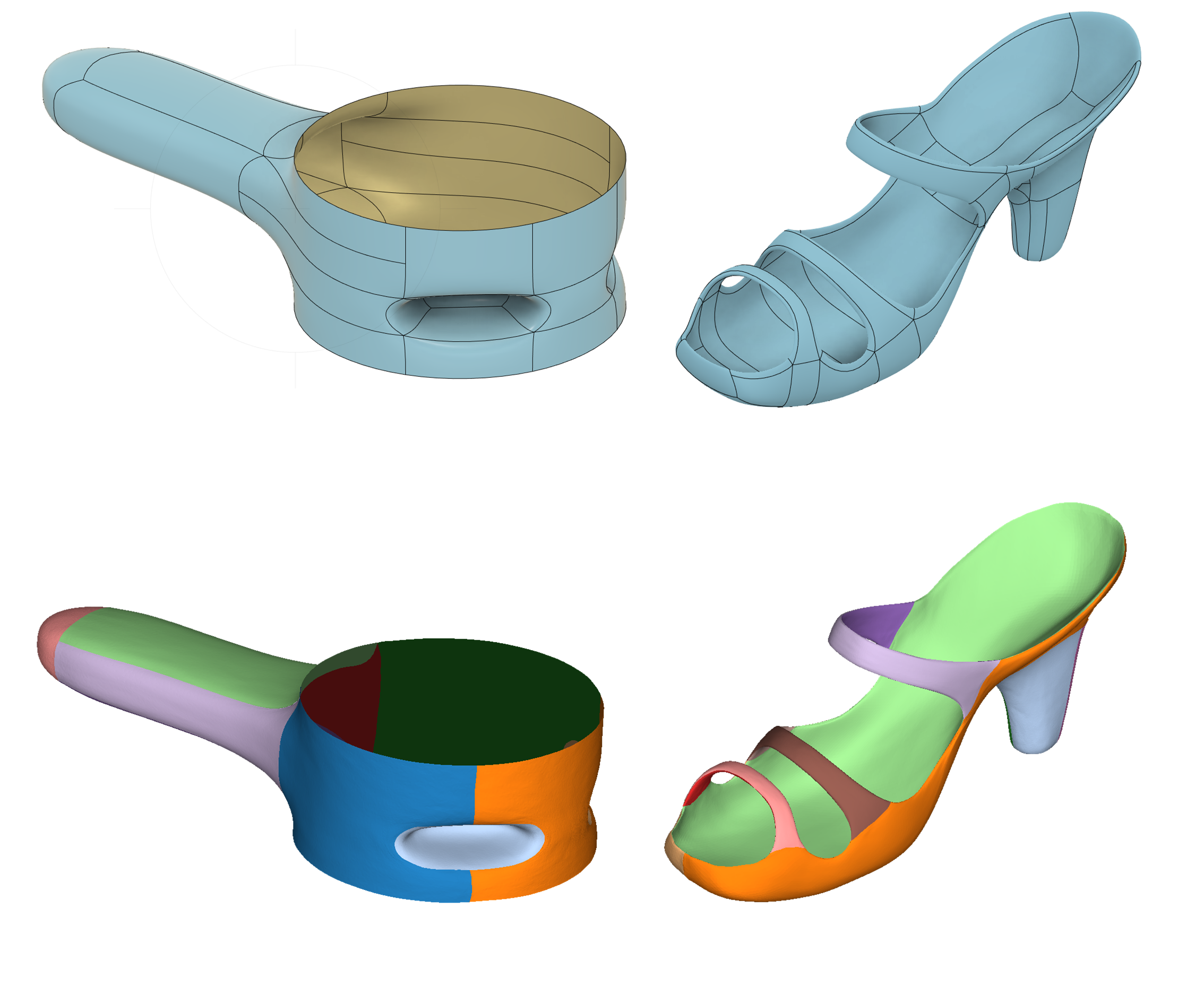}
    \put(30,45){(a)}
    \put(75,45){(b)}
    \put(30,3){(c)}
    \put(75,3){(d)}
    \end{overpic}
    \caption{\textbf{Comparisons between the results produced by T-splines (top row) and by our method (bottom row)}.  The T-spine surfaces of the \textit{Catcher} model (a) and the \textit{Shoe} model (b) contain 32 and 76 patches, respectively. Our neural parametric surfaces can represent the \textit{Catcher} (c) with 10 patches and the \textit{Shoe} (d) with 13 patches.}
    \label{fig:comp_w_tspline2}
\end{figure}

\subsubsection{Comparison with traditional parametric representations}

We discuss the difference between our representation and the traditional parametric representations. In particular, we focus on the Hermite Coons patches that are defined on a \textit{rectangle} domain, and T-spline surfaces that support T-junctures and local refinement.

We show the results of different methods in Fig.~\ref{fig:comp_w_coons}. On the left part, we show the comparison between Hermite Coons patches and our result. We can see that our representation can better model the wrinkles and the overall shape of the skirt. Hermite Coons patches take as input the partial derivatives on the boundary curve to produce the shape, hence the wrinkles are extended from the boundary information, while incapable of capturing the details of the underlying geometry.

T-spline surfaces are flexible as they can accommodate T-junctures and thus the local refinement need not go through the entire parametric domain. We show additional comparisons in Fig.~\ref{fig:comp_w_tspline2} where our method can achieve even more \textit{flexible} patch partitioning of the given surface, using fewer (10 and 13) patches to fit respective T-spline surfaces of \textit{Catcher} with 32 surface patches and \textit{Shoe} with 76 surface patches, and at the same time retain \textit{faithful} approximations of the surfaces modeled by the T-spline surfaces. Each color indicates a surface patch in our neural parametric surface.

\subsection{Ablation study}
\label{sec:ablation}

\begin{figure}
    \centering
    \begin{overpic}
        [width=\linewidth, trim = 0 0 500 0]{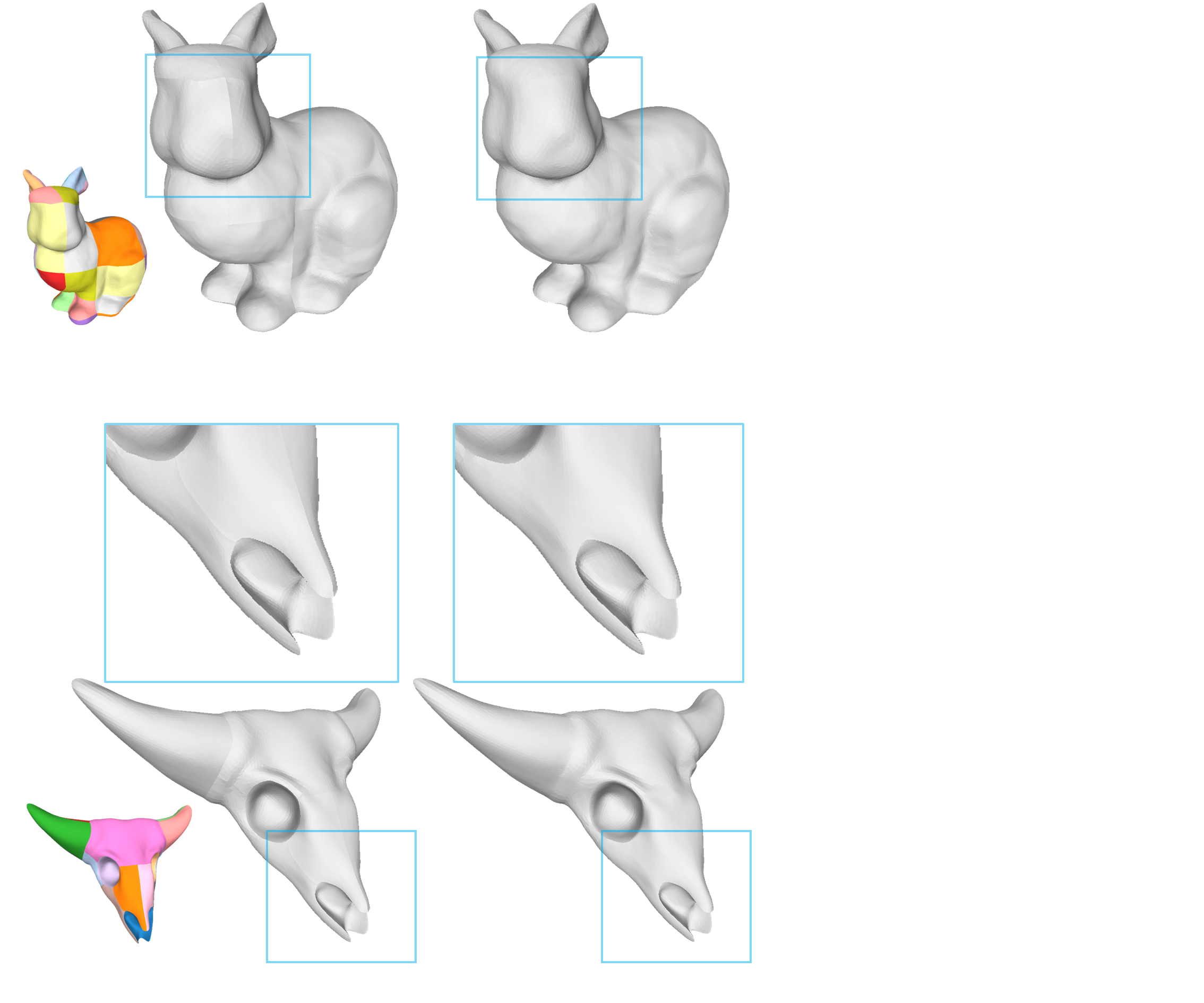}
        \put(0,62){(a) Layout}
        \put(20,62){(b) w/o $L_{\mathrm{smooth}}$}
        \put(53,62){(c) with $L_{\mathrm{smooth}}$}
        \put(0,0){(d) Layout}
        \put(20,0){(e) w/o $L_{\mathrm{smooth}}$}
        \put(53,0){(f) with $L_{\mathrm{smooth}}$}
    \end{overpic}
    \caption{\textbf{Ablation on the loss term $L_{\mathrm{smooth}}$ (Eq.~\ref{eq:g1_constraint}).} 
    Partitionings of given shapes are given in (a) and (d).
    Without $L_{\mathrm{smooth}}$, discontinuous joins between two adjacent patches are observed (see the enclosed regions in (b, e)).
    Incorporating $L_{\mathrm{smooth}}$ improves the normal continuity across the patch boundary, and thus approximates $G^1$-continuity along the boundary (c, f).}
    \label{fig:ablation_g1}
\end{figure}

\textbf{Enforcing smoothness across adjacent patches. }
We examine the effects of incorporation of the smoothness term (Eq.~\ref{eq:g1_constraint}) on shared boundaries between adjacent patches, as shown in Fig.~\ref{fig:ablation_g1}. To identify which portions of a shared boundary curve are smooth, we calculate the dihedral angles for each mesh edge along that boundary. When the dihedral angle exceeds a set threshold (in this work, $\pi/4$), the corresponding section is deemed smooth; otherwise, it is considered sharp. We aggregate smooth edges from all shared boundaries and apply the smoothness term. Our results in Fig.~\ref{fig:ablation_g1} demonstrate that this simple technique effectively enforces empirical $G^1$ continuity between adjacent patches. Although our method can only guarantee $G^0$-continuity along the patch boundary, the addition of the smoothness term (Eq.~\ref{eq:g1_constraint}) yields continuous normal transition when needed and hence, closely approximates $G^1$ continuity.

\begin{figure}[h!tb]
    \centering
    \begin{overpic}
    [width=\linewidth, trim = 0 350 0 0]{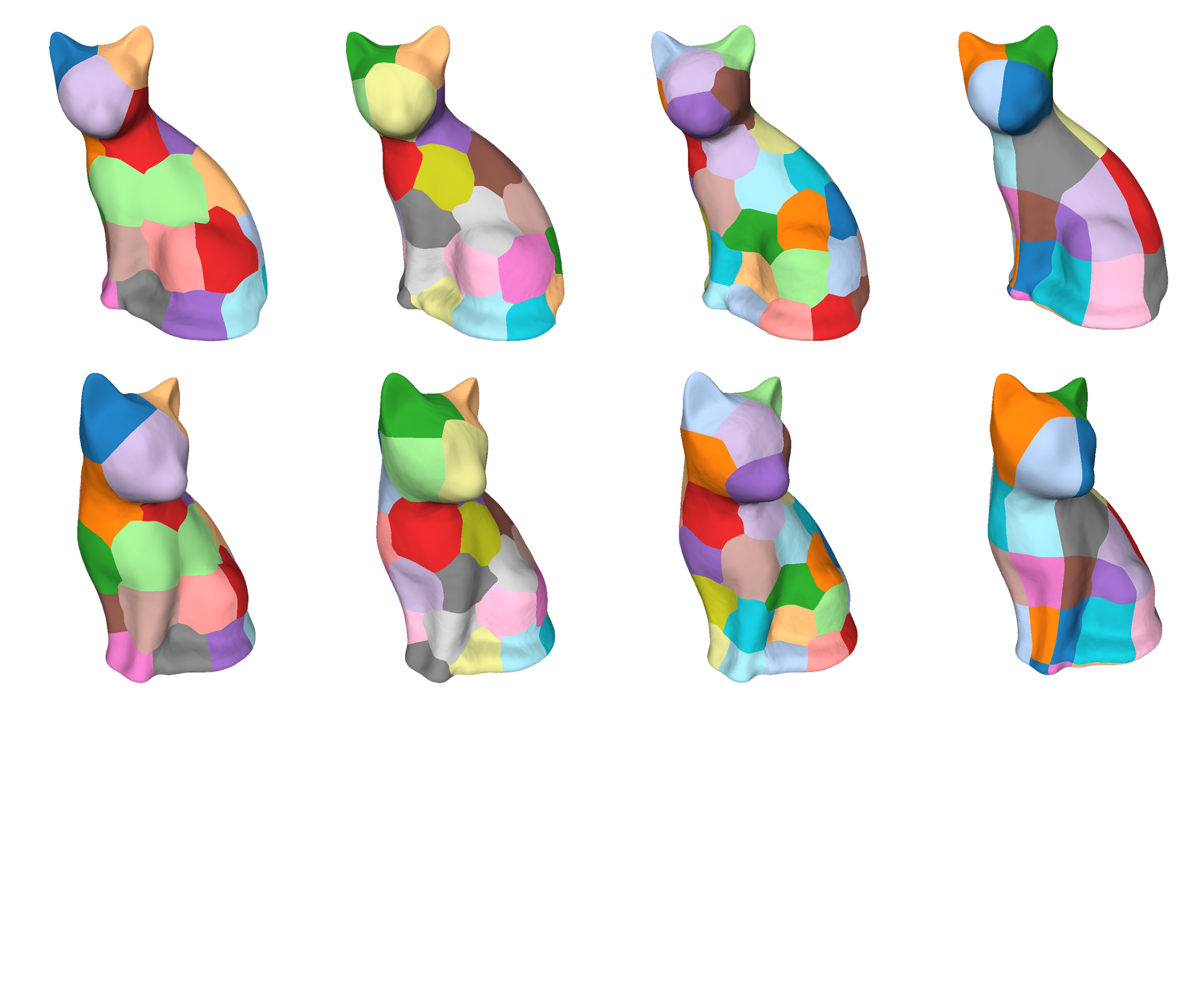}
    \put(5,0){(a) VSA30}
    \put(32,0){(b) VSA40}
    \put(56,0){(c) VSA50}
    \put(78,0){(d) LoopyCuts}
    \end{overpic}
    \caption{\textbf{Robustness of our method to various input patch layouts.} The fitting results of the \textit{Cat} model are produced with different patch layouts. (a), (b), and (c) use the patch layouts automatically generated by VSA~\cite{cohen2004variational} with 30, 40, and 50 patches for the fitting. (d) adopts the patch layout generated by LoopyCuts~\cite{livesu2020loopycuts}. 
    Visually, different fitting results do not vary much, showing the robustness of our method to various input patch layouts regarding the same shape.}
    \label{fig:robust_to_segmentation}
\end{figure}

\textbf{Robustness to different patch layouts. }
We examine the impacts of different patch layouts on the fitting performance of our Neural Parametric Surface representation. 
Specifically, on the \textit{Cat} model, we compare a patch layout generated by LoopyCut (with 46 patches) to three patch layouts generated by VSA~\cite{cohen2004variational} (each containing 30, 40, or 50 patches, respectively). This comparison, quantitatively summarized in Tab.~\ref{tab:robust_to_segmentation} and qualitatively illustrated in Fig.~\ref{fig:robust_to_segmentation}, shows that our method is relatively insensitive to the patch layout and can produce fitting results with similar performance. We observed a slight performance improvement in the P2S metric when using layouts with more patches, specifically using the one generated by LoopyCuts or the 50-patch layout by VSA. Conversely, the Hausdorff distance (HD) shows negligible variance based on the number of patches. In terms of Normal Angular Error (NAE), the layout generated by LoopyCut slightly outperforms those created by VSA.

\begin{table}[]
    \centering
    \caption{Our Neural Parametric Surface can fit the same shape robustly given different patch layouts. $|\mathcal{S}|$ is the number of patches in the given layout.}
    \begin{tabular}{c|c c c c}
    \hline
        & $|\mathcal{S}|$ & P2S & HD & NAE \\
        & & $\times 10^{-3}$ & $\times 10^{-3}$ & (deg) \\
    \hline
    VSA & 30 & 0.793 & 6.981 & 4.93 \\
        & 40 & 0.703 & 5.982 & 4.93 \\
        & 50 & 0.667 & 5.992 & 4.80 \\
    \hline
    LoopyCuts & 46 & 0.619 & 6.450 & 4.18 \\
    \hline
    \end{tabular}
    \label{tab:robust_to_segmentation}
\end{table}

\begin{figure}[h!tb]
    \centering
    \begin{overpic}
    [width=\linewidth, trim = 0 780 600 0]{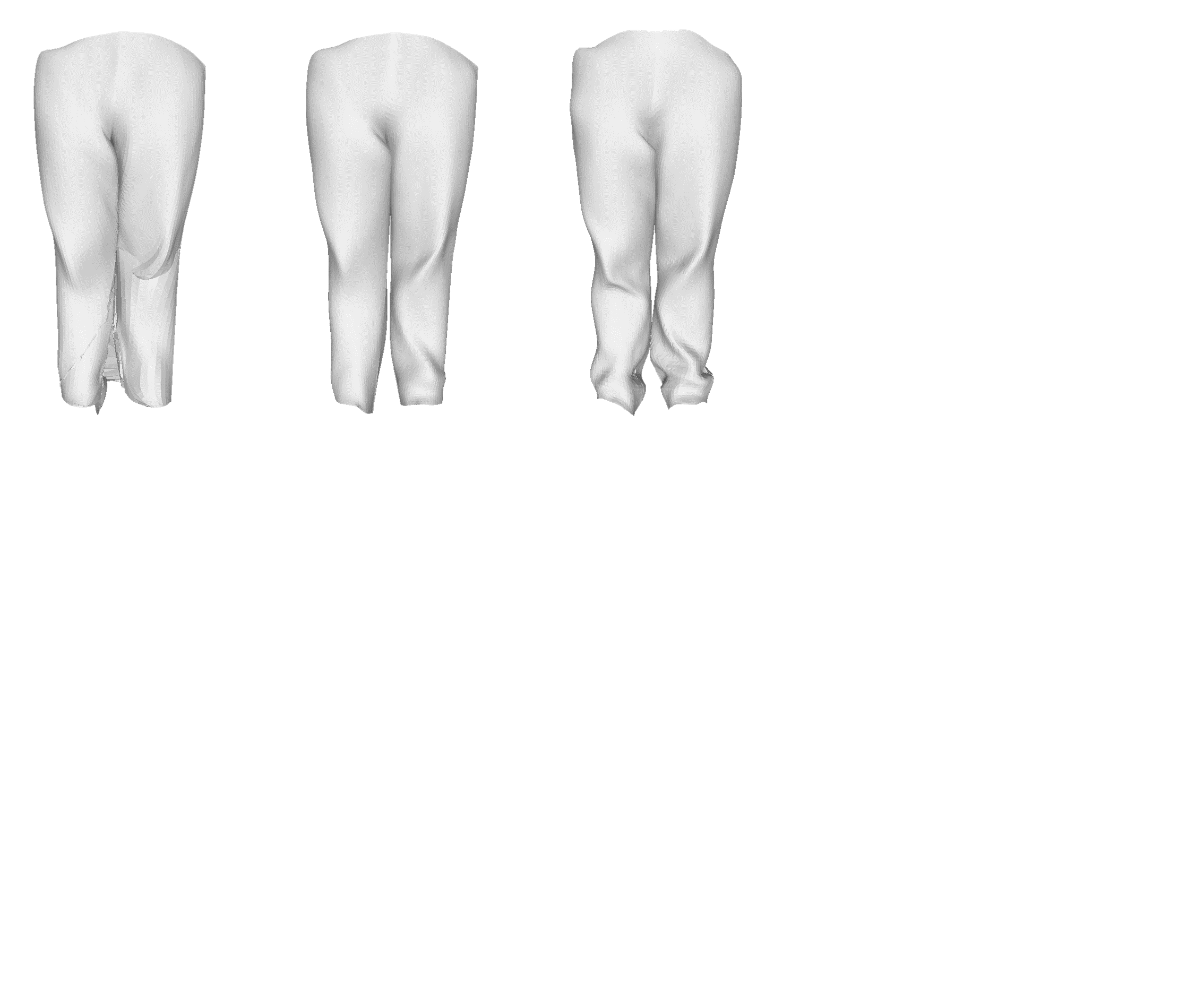}
    \put(0,0){(a) $D=3$ learnable}
    \put(40,0){(b) $D=3$ fixed}
    \put(75,0){(c) $D=128$}
    \end{overpic}
    \caption{
    \textbf{Ablation on the different dimensions of learnable space to embed the feature complex.} Compared with the setting $D=3$ learnable or $D=3$ fixed, our design choice (right) provides more degrees of freedom and thus enhances the overall expressiveness of the neural parametric surfaces, allowing them to effectively reconstruct the wrinkles.
    }
    \label{fig:different_D}
\end{figure}

\textbf{Dimension of the embedding space for feature complex.}
Through experiments, we observed that networks using a low dimensional feature space ($D=3$) cannot faithfully fit the given shape. We consider two settings: 1) the 3-dimensional embedding space is learnable as did throughout the paper, and 2) using the anchor points $\mathbf{p}_k$ in Eq.~\ref{eq:node_fitting} as the vertices of the complex and thus embed it in 3D. In the first scenario, the network model yields self-intersected patches, struggling to fit the target surface, as shown in Fig.~\ref{fig:different_D}. In the second scenario, since the anchor points are given as the vertices of the complex, the network model can produce a better approximation of the target shape than the previous setting. However, due to the lack of extra freedom, the resulting surface patches fail to reproduce highly curved wrinkles, as compared to our design choice (right). Incorporating the anchor points ($\mathbf{p}_k$) as part of the feature complex seems a feasible choice, however, note that this design will limit the proposed representation to \textit{only} the surface fitting application where anchor points are provided \textit{a priori}. For applications like linearly interpolating the latent shape space or reconstructing imperfect data, the anchor points are \textit{unknown}; interpolating the anchors linearly between two key-frame shapes can incur undesired artifacts. Therefore, it is more desirable that the feature complex can be learned through optimization.

\textbf{Network design.} 
We employ an MLP network with $L=12$ layers, each having $N=256$ neurons. The feature complex is embedded in a $D$-dimensional space, where $D=128$. Here we evaluate these design choices (denoted as the \textit{default setting}) via an ablation study. We compare the default setting to a setting with fewer layers ($L=6$) and four settings with different feature space dimensions ($D=\{32, 64\}$). The results are presented in Table~\ref{tab:ablation}. Our ablation study involves two shapes (\textit{Fandisk}, a CAD model, and \textit{Pants}, a free-form model). We observed that different settings can obtain similar results (both in terms of the geometry and of the performance metrics) on \textit{Fandisk} that consists of smooth surface patches. On the other hand, we can see that a network with high capacity (with a larger $D$ or $L$) can yield better results in terms of fitting errors on \textit{Pants}. We also tested with a wider network with $N=512$. While the results (again both geometry-wise and metric-wise) are similar, it requires more than 20 MB to store the trained network. For a consistent experimental setting, we, therefore, chose to use the current setting to produce all results throughout the paper.

\begin{table}[]
    \centering
    \caption{\textbf{Ablation study on network design.} Our default setting uses an MLP network with 12 layers ($L$), each having 256 neurons ($N$), and the dimension of the feature space is 128 ($D$). The performance metrics show that the differences on \textit{Fandisk}, a CAD model, are marginal. However, better performance is attained with a more powerful setting on the \textit{Pants}, a free-form shape with highly curved features. \vspace{-5pt}
    }
    \scalebox{0.9}{
    \begin{tabular}{c c|c c c c }
        \hline 
         &  & P2S & HD & NAE & NN \\
         Method & Shape &  $\times 10^{-3}$ & $\times 10^{-3}$ & (deg) & Storage \\
        \hline
        \multirow{2}{*}{$D = 32$}
            & Fandisk & 0.585 & 7.990 & 1.98 & 5.4 MB \\
            & Pants & 1.520 & 27.38 & 7.56 & 5.4 MB \\
        \hline
        \multirow{2}{*}{$D = 64$}
            & Fandisk & 0.832 & 11.711 & 2.14 & 5.4 MB \\
            & Pants & 1.470 & 20.433 & 7.68 & 5.4 MB \\
        \hline
        \multirow{2}{*}{$L = 6$}
            & Fandisk & 0.523 & 9.586 & 2.21 & 2.2 MB \\
            & Pants & 1.360 & 25.231 & 7.57 & 2.2 MB \\
        \hline
        \multirow{2}{*}{Default}
            & Fandisk & 0.572 & 8.743 & 2.06 & 5.5 MB \\
            & Pants & 1.351 & 18.269 & 7.36 & 5.5 MB \\
        \hline
    \end{tabular}}
    \label{tab:ablation}
\end{table}

\subsection{Limitations}

First, our algorithm requires the input shape to come with a partitioning. Although our design can handle surfaces with various layouts and can flexibly handle $n$-sided topological patches, it is currently unable to handle patches with holes. We now process patches with holes by further partitioning them into simpler sub-patches. We will explore other strategies to overcome this limitation so that a wider range of automatic segmentation tools (e.g., semantic segmentation) can be directly used to generate a layout of a shape for our pipeline. 

Second, in terms of model sizes, although like other neural representations, our neural parametric surface is more compact than point clouds and meshes, its storage requirements are higher than conventional parametric representations such as splines. The enhanced modeling flexibility comes at the expense of a larger number of (network) parameters. Therefore, an interesting direction to explore is the development of a more compact network representation to reduce the model size of the neural parametric surface. We will explore network pruning or other optimization techniques in the near future to simplify the network architecture or sparsify network parameters, without compromising model accuracy. It is also worth noting that the neural parametric surface is a compact representation of a shape space because we can use just one network to encode and decode all the shapes (10K hands or 4500 garments) within this collection.

Third, the training and optimization process for a neural parametric surface currently takes approximately 5-30 minutes. Improving the efficiency of shape fitting and reconstruction would be highly beneficial, and it could open up new possibilities for applications like interactive design and shape editing. We will explore possible schemes to accelerate the training process or pre-train the model, to support Neural Parametric Surfaces to be updated at an interactive rate.

Finally, for highly concave surface patches, parameterizing them on a $n$-sided convex polygon might occasionally produce foldover triangles, although this chance is low due to the powerful capabilities of deep neural networks. We observed a few instances of self-intersecting triangles in the concave region of the top patch (highlighted in brown) of the \textit{Fandisk}; see Fig.~\ref{fig:comparison} for the patch in Fandisk. To mitigate this issue, we can segment the concave region into multiple convex ones, which helps reduce the occurrence of foldover triangles. However, it is worth exploring a more systematic approach, for example, to allow the corners of the polygonal region to be optimized during training, or to employ more advanced interpolation schemes that can handle concave regions effectively. 

\section{Conclusions}\label{sec:conclusion}

We presented \textit{Neural Parametric Surface}, the first piecewise parametric surface representation equipped with deep neural networks. The key components are a learnable feature complex and a shared mapping function implemented as an MLP network. Neural Parametric Surfaces extend the range of shapes that can be effectively modeled as learned representations, and provide a simplified means of representing complex models with continuous and highly flexible patch layouts, compared to prior parametric approaches. This representation can also be used to learn a latent space from a collection of shapes for different downstream tasks. We have demonstrated its usefulness for shape interpolation, editing, and reconstruction from limited or noisy data.

\bibliographystyle{ACM-Reference-Format}
\bibliography{ref}


\begin{thebibliography}{47}


\ifx \showCODEN    \undefined \def \showCODEN     #1{\unskip}     \fi
\ifx \showDOI      \undefined \def \showDOI       #1{#1}\fi
\ifx \showISBNx    \undefined \def \showISBNx     #1{\unskip}     \fi
\ifx \showISBNxiii \undefined \def \showISBNxiii  #1{\unskip}     \fi
\ifx \showISSN     \undefined \def \showISSN      #1{\unskip}     \fi
\ifx \showLCCN     \undefined \def \showLCCN      #1{\unskip}     \fi
\ifx \shownote     \undefined \def \shownote      #1{#1}          \fi
\ifx \showarticletitle \undefined \def \showarticletitle #1{#1}   \fi
\ifx \showURL      \undefined \def \showURL       {\relax}        \fi
\providecommand\bibfield[2]{#2}
\providecommand\bibinfo[2]{#2}
\providecommand\natexlab[1]{#1}
\providecommand\showeprint[2][]{arXiv:#2}

\bibitem[Aigerman et~al\mbox{.}(2022)]%
        {aigerman2022neural}
\bibfield{author}{\bibinfo{person}{Noam Aigerman}, \bibinfo{person}{Kunal Gupta}, \bibinfo{person}{Vladimir~G Kim}, \bibinfo{person}{Siddhartha Chaudhuri}, \bibinfo{person}{Jun Saito}, {and} \bibinfo{person}{Thibault Groueix}.} \bibinfo{year}{2022}\natexlab{}.
\newblock \showarticletitle{Neural Jacobian Fields: Learning Intrinsic Mappings of Arbitrary Meshes}.
\newblock \bibinfo{journal}{\emph{arXiv preprint arXiv:2205.02904}} (\bibinfo{year}{2022}).
\newblock


\bibitem[Bae et~al\mbox{.}(2008)]%
        {bae2008ilovesketch}
\bibfield{author}{\bibinfo{person}{Seok-Hyung Bae}, \bibinfo{person}{Ravin Balakrishnan}, {and} \bibinfo{person}{Karan Singh}.} \bibinfo{year}{2008}\natexlab{}.
\newblock \showarticletitle{ILoveSketch: as-natural-as-possible sketching system for creating 3d curve models}. In \bibinfo{booktitle}{\emph{Proceedings of the 21st annual ACM symposium on User interface software and technology}}. \bibinfo{pages}{151--160}.
\newblock


\bibitem[Bedna{\v{r}}{\'\i}k et~al\mbox{.}(2020)]%
        {bednarik2020shape}
\bibfield{author}{\bibinfo{person}{Jan Bedna{\v{r}}{\'\i}k}, \bibinfo{person}{Shaifali Parashar}, \bibinfo{person}{Erhan Gundogdu}, \bibinfo{person}{Mathieu Salzmann}, {and} \bibinfo{person}{Pascal Fua}.} \bibinfo{year}{2020}\natexlab{}.
\newblock \showarticletitle{Shape reconstruction by learning differentiable surface representations}. In \bibinfo{booktitle}{\emph{Proceedings of the IEEE/CVF Conference on Computer Vision and Pattern Recognition}}. \bibinfo{pages}{4716--4725}.
\newblock


\bibitem[Bhatnagar et~al\mbox{.}(2019)]%
        {bhatnagar2019multi}
\bibfield{author}{\bibinfo{person}{Bharat~Lal Bhatnagar}, \bibinfo{person}{Garvita Tiwari}, \bibinfo{person}{Christian Theobalt}, {and} \bibinfo{person}{Gerard Pons-Moll}.} \bibinfo{year}{2019}\natexlab{}.
\newblock \showarticletitle{Multi-garment net: Learning to dress 3d people from images}. In \bibinfo{booktitle}{\emph{proceedings of the IEEE/CVF international conference on computer vision}}. \bibinfo{pages}{5420--5430}.
\newblock


\bibitem[Born et~al\mbox{.}(2021)]%
        {born2021layout}
\bibfield{author}{\bibinfo{person}{Janis Born}, \bibinfo{person}{Patrick Schmidt}, {and} \bibinfo{person}{Leif Kobbelt}.} \bibinfo{year}{2021}\natexlab{}.
\newblock \showarticletitle{Layout embedding via combinatorial optimization}. In \bibinfo{booktitle}{\emph{Computer Graphics Forum}}, Vol.~\bibinfo{volume}{40}. Wiley Online Library, \bibinfo{pages}{277--290}.
\newblock


\bibitem[Campen(2017)]%
        {campen2017partitioning}
\bibfield{author}{\bibinfo{person}{Marcel Campen}.} \bibinfo{year}{2017}\natexlab{}.
\newblock \showarticletitle{Partitioning surfaces into quadrilateral patches: A survey}. In \bibinfo{booktitle}{\emph{Computer Graphics Forum}}, Vol.~\bibinfo{volume}{36}. Wiley Online Library, \bibinfo{pages}{567--588}.
\newblock


\bibitem[Chibane et~al\mbox{.}(2020)]%
        {chibane2020neural}
\bibfield{author}{\bibinfo{person}{Julian Chibane}, \bibinfo{person}{Gerard Pons-Moll}, {et~al\mbox{.}}} \bibinfo{year}{2020}\natexlab{}.
\newblock \showarticletitle{Neural unsigned distance fields for implicit function learning}.
\newblock \bibinfo{journal}{\emph{Advances in Neural Information Processing Systems}}  \bibinfo{volume}{33} (\bibinfo{year}{2020}), \bibinfo{pages}{21638--21652}.
\newblock


\bibitem[Cohen-Steiner et~al\mbox{.}(2004)]%
        {cohen2004variational}
\bibfield{author}{\bibinfo{person}{David Cohen-Steiner}, \bibinfo{person}{Pierre Alliez}, {and} \bibinfo{person}{Mathieu Desbrun}.} \bibinfo{year}{2004}\natexlab{}.
\newblock \showarticletitle{Variational shape approximation}.
\newblock In \bibinfo{booktitle}{\emph{ACM SIGGRAPH 2004 Papers}}. \bibinfo{pages}{905--914}.
\newblock


\bibitem[Deng et~al\mbox{.}(2021)]%
        {deng2021deformed}
\bibfield{author}{\bibinfo{person}{Yu Deng}, \bibinfo{person}{Jiaolong Yang}, {and} \bibinfo{person}{Xin Tong}.} \bibinfo{year}{2021}\natexlab{}.
\newblock \showarticletitle{Deformed implicit field: Modeling 3d shapes with learned dense correspondence}. In \bibinfo{booktitle}{\emph{Proceedings of the IEEE/CVF Conference on Computer Vision and Pattern Recognition}}. \bibinfo{pages}{10286--10296}.
\newblock


\bibitem[Deng et~al\mbox{.}(2020)]%
        {deng2020better}
\bibfield{author}{\bibinfo{person}{Zhantao Deng}, \bibinfo{person}{Jan Bedna{\v{r}}{\'\i}k}, \bibinfo{person}{Mathieu Salzmann}, {and} \bibinfo{person}{Pascal Fua}.} \bibinfo{year}{2020}\natexlab{}.
\newblock \showarticletitle{Better patch stitching for parametric surface reconstruction}. In \bibinfo{booktitle}{\emph{2020 International Conference on 3D Vision (3DV)}}. IEEE, \bibinfo{pages}{593--602}.
\newblock


\bibitem[Deprelle et~al\mbox{.}(2022)]%
        {deprelle2022learning}
\bibfield{author}{\bibinfo{person}{Theo Deprelle}, \bibinfo{person}{Thibault Groueix}, \bibinfo{person}{Noam Aigerman}, \bibinfo{person}{Vladimir~G Kim}, {and} \bibinfo{person}{Mathieu Aubry}.} \bibinfo{year}{2022}\natexlab{}.
\newblock \showarticletitle{Learning Joint Surface Atlases}.
\newblock \bibinfo{journal}{\emph{arXiv preprint arXiv:2206.06273}} (\bibinfo{year}{2022}).
\newblock


\bibitem[Deprelle et~al\mbox{.}(2019)]%
        {deprelle2019learning}
\bibfield{author}{\bibinfo{person}{Theo Deprelle}, \bibinfo{person}{Thibault Groueix}, \bibinfo{person}{Matthew Fisher}, \bibinfo{person}{Vladimir Kim}, \bibinfo{person}{Bryan Russell}, {and} \bibinfo{person}{Mathieu Aubry}.} \bibinfo{year}{2019}\natexlab{}.
\newblock \showarticletitle{Learning elementary structures for 3d shape generation and matching}.
\newblock \bibinfo{journal}{\emph{Advances in Neural Information Processing Systems}}  \bibinfo{volume}{32} (\bibinfo{year}{2019}).
\newblock


\bibitem[Floater(2003)]%
        {floater2003mean}
\bibfield{author}{\bibinfo{person}{Michael~S Floater}.} \bibinfo{year}{2003}\natexlab{}.
\newblock \showarticletitle{Mean value coordinates}.
\newblock \bibinfo{journal}{\emph{Computer aided geometric design}} \bibinfo{volume}{20}, \bibinfo{number}{1} (\bibinfo{year}{2003}), \bibinfo{pages}{19--27}.
\newblock


\bibitem[Gadelha et~al\mbox{.}(2021)]%
        {gadelha2021deep}
\bibfield{author}{\bibinfo{person}{Matheus Gadelha}, \bibinfo{person}{Rui Wang}, {and} \bibinfo{person}{Subhransu Maji}.} \bibinfo{year}{2021}\natexlab{}.
\newblock \showarticletitle{Deep manifold prior}. In \bibinfo{booktitle}{\emph{Proceedings of the IEEE/CVF International Conference on Computer Vision}}. \bibinfo{pages}{1107--1116}.
\newblock


\bibitem[Gao et~al\mbox{.}(2022)]%
        {gao2022dart}
\bibfield{author}{\bibinfo{person}{Daiheng Gao}, \bibinfo{person}{Yuliang Xiu}, \bibinfo{person}{Kailin Li}, \bibinfo{person}{Lixin Yang}, \bibinfo{person}{Feng Wang}, \bibinfo{person}{Peng Zhang}, \bibinfo{person}{Bang Zhang}, \bibinfo{person}{Cewu Lu}, {and} \bibinfo{person}{Ping Tan}.} \bibinfo{year}{2022}\natexlab{}.
\newblock \showarticletitle{{DART: Articulated Hand Model with Diverse Accessories and Rich Textures}}. In \bibinfo{booktitle}{\emph{Thirty-sixth Conference on Neural Information Processing Systems Datasets and Benchmarks Track}}.
\newblock


\bibitem[Gropp et~al\mbox{.}(2020)]%
        {gropp2020implicit}
\bibfield{author}{\bibinfo{person}{Amos Gropp}, \bibinfo{person}{Lior Yariv}, \bibinfo{person}{Niv Haim}, \bibinfo{person}{Matan Atzmon}, {and} \bibinfo{person}{Yaron Lipman}.} \bibinfo{year}{2020}\natexlab{}.
\newblock \showarticletitle{Implicit Geometric Regularization for Learning Shapes}. In \bibinfo{booktitle}{\emph{International Conference on Machine Learning}}. PMLR, \bibinfo{pages}{3789--3799}.
\newblock


\bibitem[Groueix et~al\mbox{.}(2018a)]%
        {groueix20183d}
\bibfield{author}{\bibinfo{person}{Thibault Groueix}, \bibinfo{person}{Matthew Fisher}, \bibinfo{person}{Vladimir~G Kim}, \bibinfo{person}{Bryan~C Russell}, {and} \bibinfo{person}{Mathieu Aubry}.} \bibinfo{year}{2018}\natexlab{a}.
\newblock \showarticletitle{3d-coded: 3d correspondences by deep deformation}. In \bibinfo{booktitle}{\emph{Proceedings of the European Conference on Computer Vision (ECCV)}}. \bibinfo{pages}{230--246}.
\newblock


\bibitem[Groueix et~al\mbox{.}(2018b)]%
        {groueix2018papier}
\bibfield{author}{\bibinfo{person}{Thibault Groueix}, \bibinfo{person}{Matthew Fisher}, \bibinfo{person}{Vladimir~G Kim}, \bibinfo{person}{Bryan~C Russell}, {and} \bibinfo{person}{Mathieu Aubry}.} \bibinfo{year}{2018}\natexlab{b}.
\newblock \showarticletitle{A papier-m{\^a}ch{\'e} approach to learning 3d surface generation}. In \bibinfo{booktitle}{\emph{Proceedings of the IEEE conference on computer vision and pattern recognition}}. \bibinfo{pages}{216--224}.
\newblock


\bibitem[Guo et~al\mbox{.}(2022b)]%
        {guo2022complexgen}
\bibfield{author}{\bibinfo{person}{Haoxiang Guo}, \bibinfo{person}{Shilin Liu}, \bibinfo{person}{Hao Pan}, \bibinfo{person}{Yang Liu}, \bibinfo{person}{Xin Tong}, {and} \bibinfo{person}{Baining Guo}.} \bibinfo{year}{2022}\natexlab{b}.
\newblock \showarticletitle{ComplexGen: CAD reconstruction by B-rep chain complex generation}.
\newblock \bibinfo{journal}{\emph{ACM Transactions on Graphics (TOG)}} \bibinfo{volume}{41}, \bibinfo{number}{4} (\bibinfo{year}{2022}), \bibinfo{pages}{1--18}.
\newblock


\bibitem[Guo et~al\mbox{.}(2022a)]%
        {guo2022implicit}
\bibfield{author}{\bibinfo{person}{Hao-Xiang Guo}, \bibinfo{person}{Yang Liu}, \bibinfo{person}{Hao Pan}, {and} \bibinfo{person}{Baining Guo}.} \bibinfo{year}{2022}\natexlab{a}.
\newblock \showarticletitle{Implicit Conversion of Manifold B-Rep Solids by Neural Halfspace Representation}.
\newblock \bibinfo{journal}{\emph{ACM Transactions on Graphics (TOG)}} \bibinfo{volume}{41}, \bibinfo{number}{6} (\bibinfo{year}{2022}), \bibinfo{pages}{1--15}.
\newblock


\bibitem[Kingma and Ba(2014)]%
        {kingma2014adam}
\bibfield{author}{\bibinfo{person}{Diederik~P Kingma} {and} \bibinfo{person}{Jimmy Ba}.} \bibinfo{year}{2014}\natexlab{}.
\newblock \showarticletitle{Adam: A method for stochastic optimization}.
\newblock \bibinfo{journal}{\emph{arXiv preprint arXiv:1412.6980}} (\bibinfo{year}{2014}).
\newblock


\bibitem[Livesu et~al\mbox{.}(2020)]%
        {livesu2020loopycuts}
\bibfield{author}{\bibinfo{person}{Marco Livesu}, \bibinfo{person}{Nico Pietroni}, \bibinfo{person}{Enrico Puppo}, \bibinfo{person}{Alla Sheffer}, {and} \bibinfo{person}{Paolo Cignoni}.} \bibinfo{year}{2020}\natexlab{}.
\newblock \showarticletitle{LoopyCuts: Practical Feature-Preserving Block Decomposition for Strongly Hex-Dominant Meshing}.
\newblock \bibinfo{journal}{\emph{ACM Trans. Graph.}} \bibinfo{volume}{39}, \bibinfo{number}{4}, Article \bibinfo{articleno}{121} (\bibinfo{date}{aug} \bibinfo{year}{2020}), \bibinfo{numpages}{17}~pages.
\newblock
\showISSN{0730-0301}
\urldef\tempurl%
\url{https://doi.org/10.1145/3386569.3392472}
\showDOI{\tempurl}


\bibitem[Lorensen and Cline(1998)]%
        {lorensen1998marching}
\bibfield{author}{\bibinfo{person}{William~E Lorensen} {and} \bibinfo{person}{Harvey~E Cline}.} \bibinfo{year}{1998}\natexlab{}.
\newblock \showarticletitle{Marching cubes: A high resolution 3D surface construction algorithm}.
\newblock In \bibinfo{booktitle}{\emph{Seminal graphics: pioneering efforts that shaped the field}}. \bibinfo{pages}{347--353}.
\newblock


\bibitem[Low and Lee(2022)]%
        {low2022minimal}
\bibfield{author}{\bibinfo{person}{Weng~Fei Low} {and} \bibinfo{person}{Gim~Hee Lee}.} \bibinfo{year}{2022}\natexlab{}.
\newblock \showarticletitle{Minimal Neural Atlas: Parameterizing Complex Surfaces with Minimal Charts and Distortion}. In \bibinfo{booktitle}{\emph{European Conference on Computer Vision}}. Springer, \bibinfo{pages}{465--481}.
\newblock


\bibitem[Martel et~al\mbox{.}(2021)]%
        {martel2021acorn}
\bibfield{author}{\bibinfo{person}{Julien~NP Martel}, \bibinfo{person}{David~B Lindell}, \bibinfo{person}{Connor~Z Lin}, \bibinfo{person}{Eric~R Chan}, \bibinfo{person}{Marco Monteiro}, {and} \bibinfo{person}{Gordon Wetzstein}.} \bibinfo{year}{2021}\natexlab{}.
\newblock \showarticletitle{Acorn: adaptive coordinate networks for neural scene representation}.
\newblock \bibinfo{journal}{\emph{ACM Transactions on Graphics (TOG)}} \bibinfo{volume}{40}, \bibinfo{number}{4} (\bibinfo{year}{2021}), \bibinfo{pages}{1--13}.
\newblock


\bibitem[Morreale et~al\mbox{.}(2022)]%
        {morreale2022neural}
\bibfield{author}{\bibinfo{person}{Luca Morreale}, \bibinfo{person}{Noam Aigerman}, \bibinfo{person}{Paul Guerrero}, \bibinfo{person}{Vladimir~G Kim}, {and} \bibinfo{person}{Niloy~J Mitra}.} \bibinfo{year}{2022}\natexlab{}.
\newblock \showarticletitle{Neural Convolutional Surfaces}. In \bibinfo{booktitle}{\emph{Proceedings of the IEEE/CVF Conference on Computer Vision and Pattern Recognition}}. \bibinfo{pages}{19333--19342}.
\newblock


\bibitem[Morreale et~al\mbox{.}(2021)]%
        {morreale2021neural}
\bibfield{author}{\bibinfo{person}{Luca Morreale}, \bibinfo{person}{Noam Aigerman}, \bibinfo{person}{Vladimir~G Kim}, {and} \bibinfo{person}{Niloy~J Mitra}.} \bibinfo{year}{2021}\natexlab{}.
\newblock \showarticletitle{Neural surface maps}. In \bibinfo{booktitle}{\emph{Proceedings of the IEEE/CVF Conference on Computer Vision and Pattern Recognition}}. \bibinfo{pages}{4639--4648}.
\newblock


\bibitem[Pan et~al\mbox{.}(2015)]%
        {pan2015flow}
\bibfield{author}{\bibinfo{person}{Hao Pan}, \bibinfo{person}{Yang Liu}, \bibinfo{person}{Alla Sheffer}, \bibinfo{person}{Nicholas Vining}, \bibinfo{person}{Chang-Jian Li}, {and} \bibinfo{person}{Wenping Wang}.} \bibinfo{year}{2015}\natexlab{}.
\newblock \showarticletitle{Flow aligned surfacing of curve networks}.
\newblock \bibinfo{journal}{\emph{ACM Transactions on Graphics (TOG)}} \bibinfo{volume}{34}, \bibinfo{number}{4} (\bibinfo{year}{2015}), \bibinfo{pages}{1--10}.
\newblock


\bibitem[Park et~al\mbox{.}(2019)]%
        {park2019deepsdf}
\bibfield{author}{\bibinfo{person}{Jeong~Joon Park}, \bibinfo{person}{Peter Florence}, \bibinfo{person}{Julian Straub}, \bibinfo{person}{Richard Newcombe}, {and} \bibinfo{person}{Steven Lovegrove}.} \bibinfo{year}{2019}\natexlab{}.
\newblock \showarticletitle{Deepsdf: Learning continuous signed distance functions for shape representation}. In \bibinfo{booktitle}{\emph{Proceedings of the IEEE/CVF conference on computer vision and pattern recognition}}. \bibinfo{pages}{165--174}.
\newblock


\bibitem[Paszke et~al\mbox{.}(2019)]%
        {NEURIPS2019_9015}
\bibfield{author}{\bibinfo{person}{Adam Paszke}, \bibinfo{person}{Sam Gross}, \bibinfo{person}{Francisco Massa}, \bibinfo{person}{Adam Lerer}, \bibinfo{person}{James Bradbury}, \bibinfo{person}{Gregory Chanan}, \bibinfo{person}{Trevor Killeen}, \bibinfo{person}{Zeming Lin}, \bibinfo{person}{Natalia Gimelshein}, \bibinfo{person}{Luca Antiga}, \bibinfo{person}{Alban Desmaison}, \bibinfo{person}{Andreas Kopf}, \bibinfo{person}{Edward Yang}, \bibinfo{person}{Zachary DeVito}, \bibinfo{person}{Martin Raison}, \bibinfo{person}{Alykhan Tejani}, \bibinfo{person}{Sasank Chilamkurthy}, \bibinfo{person}{Benoit Steiner}, \bibinfo{person}{Lu Fang}, \bibinfo{person}{Junjie Bai}, {and} \bibinfo{person}{Soumith Chintala}.} \bibinfo{year}{2019}\natexlab{}.
\newblock \showarticletitle{PyTorch: An Imperative Style, High-Performance Deep Learning Library}.
\newblock In \bibinfo{booktitle}{\emph{Advances in Neural Information Processing Systems 32}}, \bibfield{editor}{\bibinfo{person}{H.~Wallach}, \bibinfo{person}{H.~Larochelle}, \bibinfo{person}{A.~Beygelzimer}, \bibinfo{person}{F.~d\textquotesingle Alch\'{e}-Buc}, \bibinfo{person}{E.~Fox}, {and} \bibinfo{person}{R.~Garnett}} (Eds.). \bibinfo{publisher}{Curran Associates, Inc.}, \bibinfo{pages}{8024--8035}.
\newblock
\urldef\tempurl%
\url{http://papers.neurips.cc/paper/9015-pytorch-an-imperative-style-high-performance-deep-learning-library.pdf}
\showURL{%
\tempurl}


\bibitem[Pietroni et~al\mbox{.}(2022)]%
        {pietroni2022computational}
\bibfield{author}{\bibinfo{person}{Nico Pietroni}, \bibinfo{person}{Corentin Dumery}, \bibinfo{person}{Raphael Falque}, \bibinfo{person}{Mark Liu}, \bibinfo{person}{Teresa Vidal-Calleja}, {and} \bibinfo{person}{Olga Sorkine-Hornung}.} \bibinfo{year}{2022}\natexlab{}.
\newblock \showarticletitle{Computational pattern making from 3D garment models}.
\newblock \bibinfo{journal}{\emph{ACM Transactions on Graphics (TOG)}} \bibinfo{volume}{41}, \bibinfo{number}{4} (\bibinfo{year}{2022}), \bibinfo{pages}{1--14}.
\newblock


\bibitem[Pokhariya et~al\mbox{.}(2022)]%
        {pokhariya2022discretization}
\bibfield{author}{\bibinfo{person}{Chandradeep Pokhariya}, \bibinfo{person}{Shanthika Naik}, \bibinfo{person}{Astitva Srivastava}, {and} \bibinfo{person}{Avinash Sharma}.} \bibinfo{year}{2022}\natexlab{}.
\newblock \showarticletitle{Discretization-Agnostic Deep Self-Supervised 3D Surface Parameterization}.
\newblock In \bibinfo{booktitle}{\emph{SIGGRAPH Asia 2022 Technical Communications}}. \bibinfo{pages}{1--4}.
\newblock


\bibitem[Qi et~al\mbox{.}(2017)]%
        {qi2017pointnet++}
\bibfield{author}{\bibinfo{person}{Charles~Ruizhongtai Qi}, \bibinfo{person}{Li Yi}, \bibinfo{person}{Hao Su}, {and} \bibinfo{person}{Leonidas~J Guibas}.} \bibinfo{year}{2017}\natexlab{}.
\newblock \showarticletitle{Pointnet++: Deep hierarchical feature learning on point sets in a metric space}.
\newblock \bibinfo{journal}{\emph{Advances in neural information processing systems}}  \bibinfo{volume}{30} (\bibinfo{year}{2017}).
\newblock


\bibitem[Sharma et~al\mbox{.}(2020)]%
        {sharma2020parsenet}
\bibfield{author}{\bibinfo{person}{Gopal Sharma}, \bibinfo{person}{Difan Liu}, \bibinfo{person}{Subhransu Maji}, \bibinfo{person}{Evangelos Kalogerakis}, \bibinfo{person}{Siddhartha Chaudhuri}, {and} \bibinfo{person}{Radom{\'\i}r M{\v{e}}ch}.} \bibinfo{year}{2020}\natexlab{}.
\newblock \showarticletitle{Parsenet: A parametric surface fitting network for 3d point clouds}. In \bibinfo{booktitle}{\emph{European Conference on Computer Vision}}. Springer, \bibinfo{pages}{261--276}.
\newblock


\bibitem[Sitzmann et~al\mbox{.}(2020)]%
        {sitzmann2020implicit}
\bibfield{author}{\bibinfo{person}{Vincent Sitzmann}, \bibinfo{person}{Julien Martel}, \bibinfo{person}{Alexander Bergman}, \bibinfo{person}{David Lindell}, {and} \bibinfo{person}{Gordon Wetzstein}.} \bibinfo{year}{2020}\natexlab{}.
\newblock \showarticletitle{Implicit neural representations with periodic activation functions}.
\newblock \bibinfo{journal}{\emph{Advances in Neural Information Processing Systems}}  \bibinfo{volume}{33} (\bibinfo{year}{2020}), \bibinfo{pages}{7462--7473}.
\newblock


\bibitem[Smirnov et~al\mbox{.}(2020)]%
        {smirnov2020learning}
\bibfield{author}{\bibinfo{person}{Dmitriy Smirnov}, \bibinfo{person}{Mikhail Bessmeltsev}, {and} \bibinfo{person}{Justin Solomon}.} \bibinfo{year}{2020}\natexlab{}.
\newblock \showarticletitle{Learning Manifold Patch-Based Representations of Man-Made Shapes}. In \bibinfo{booktitle}{\emph{International Conference on Learning Representations}}.
\newblock


\bibitem[Takikawa et~al\mbox{.}(2021)]%
        {takikawa2021neural}
\bibfield{author}{\bibinfo{person}{Towaki Takikawa}, \bibinfo{person}{Joey Litalien}, \bibinfo{person}{Kangxue Yin}, \bibinfo{person}{Karsten Kreis}, \bibinfo{person}{Charles Loop}, \bibinfo{person}{Derek Nowrouzezahrai}, \bibinfo{person}{Alec Jacobson}, \bibinfo{person}{Morgan McGuire}, {and} \bibinfo{person}{Sanja Fidler}.} \bibinfo{year}{2021}\natexlab{}.
\newblock \showarticletitle{Neural geometric level of detail: Real-time rendering with implicit 3D shapes}. In \bibinfo{booktitle}{\emph{Proceedings of the IEEE/CVF Conference on Computer Vision and Pattern Recognition}}. \bibinfo{pages}{11358--11367}.
\newblock


\bibitem[Tarini et~al\mbox{.}(2004)]%
        {tarini2004polycube}
\bibfield{author}{\bibinfo{person}{Marco Tarini}, \bibinfo{person}{Kai Hormann}, \bibinfo{person}{Paolo Cignoni}, {and} \bibinfo{person}{Claudio Montani}.} \bibinfo{year}{2004}\natexlab{}.
\newblock \showarticletitle{Polycube-maps}.
\newblock \bibinfo{journal}{\emph{ACM transactions on graphics (TOG)}} \bibinfo{volume}{23}, \bibinfo{number}{3} (\bibinfo{year}{2004}), \bibinfo{pages}{853--860}.
\newblock


\bibitem[Tretschk et~al\mbox{.}(2020)]%
        {tretschk2020patchnets}
\bibfield{author}{\bibinfo{person}{Edgar Tretschk}, \bibinfo{person}{Ayush Tewari}, \bibinfo{person}{Vladislav Golyanik}, \bibinfo{person}{Michael Zollh{\"o}fer}, \bibinfo{person}{Carsten Stoll}, {and} \bibinfo{person}{Christian Theobalt}.} \bibinfo{year}{2020}\natexlab{}.
\newblock \showarticletitle{Patchnets: Patch-based generalizable deep implicit 3d shape representations}. In \bibinfo{booktitle}{\emph{Computer Vision--ECCV 2020: 16th European Conference, Glasgow, UK, August 23--28, 2020, Proceedings, Part XVI 16}}. Springer, \bibinfo{pages}{293--309}.
\newblock


\bibitem[Wang et~al\mbox{.}(2018)]%
        {wang2018learning}
\bibfield{author}{\bibinfo{person}{Tuanfeng~Y Wang}, \bibinfo{person}{Duygu Ceylan}, \bibinfo{person}{Jovan Popovi{\'c}}, {and} \bibinfo{person}{Niloy~J Mitra}.} \bibinfo{year}{2018}\natexlab{}.
\newblock \showarticletitle{Learning a shared shape space for multimodal garment design}.
\newblock \bibinfo{journal}{\emph{ACM Transactions on Graphics (TOG)}} \bibinfo{volume}{37}, \bibinfo{number}{6} (\bibinfo{year}{2018}), \bibinfo{pages}{1--13}.
\newblock


\bibitem[Watters et~al\mbox{.}(2019)]%
        {watters2019spatial}
\bibfield{author}{\bibinfo{person}{Nicholas Watters}, \bibinfo{person}{Loic Matthey}, \bibinfo{person}{Christopher~P Burgess}, {and} \bibinfo{person}{Alexander Lerchner}.} \bibinfo{year}{2019}\natexlab{}.
\newblock \showarticletitle{Spatial broadcast decoder: A simple architecture for learning disentangled representations in vaes}.
\newblock \bibinfo{journal}{\emph{arXiv preprint arXiv:1901.07017}} (\bibinfo{year}{2019}).
\newblock


\bibitem[Williams et~al\mbox{.}(2019)]%
        {williams2019deep}
\bibfield{author}{\bibinfo{person}{Francis Williams}, \bibinfo{person}{Teseo Schneider}, \bibinfo{person}{Claudio Silva}, \bibinfo{person}{Denis Zorin}, \bibinfo{person}{Joan Bruna}, {and} \bibinfo{person}{Daniele Panozzo}.} \bibinfo{year}{2019}\natexlab{}.
\newblock \showarticletitle{Deep geometric prior for surface reconstruction}. In \bibinfo{booktitle}{\emph{Proceedings of the IEEE/CVF Conference on Computer Vision and Pattern Recognition}}. \bibinfo{pages}{10130--10139}.
\newblock


\bibitem[Yang et~al\mbox{.}(2021)]%
        {yang2021geometry}
\bibfield{author}{\bibinfo{person}{Guandao Yang}, \bibinfo{person}{Serge Belongie}, \bibinfo{person}{Bharath Hariharan}, {and} \bibinfo{person}{Vladlen Koltun}.} \bibinfo{year}{2021}\natexlab{}.
\newblock \showarticletitle{Geometry processing with neural fields}.
\newblock \bibinfo{journal}{\emph{Advances in Neural Information Processing Systems}}  \bibinfo{volume}{34} (\bibinfo{year}{2021}), \bibinfo{pages}{22483--22497}.
\newblock


\bibitem[Yang et~al\mbox{.}(2018)]%
        {yang2018foldingnet}
\bibfield{author}{\bibinfo{person}{Yaoqing Yang}, \bibinfo{person}{Chen Feng}, \bibinfo{person}{Yiru Shen}, {and} \bibinfo{person}{Dong Tian}.} \bibinfo{year}{2018}\natexlab{}.
\newblock \showarticletitle{Foldingnet: Point cloud auto-encoder via deep grid deformation}. In \bibinfo{booktitle}{\emph{Proceedings of the IEEE conference on computer vision and pattern recognition}}. \bibinfo{pages}{206--215}.
\newblock


\bibitem[Yu et~al\mbox{.}(2022)]%
        {yu2022piecewise}
\bibfield{author}{\bibinfo{person}{Emilie Yu}, \bibinfo{person}{Rahul Arora}, \bibinfo{person}{J~Andreas Baerentzen}, \bibinfo{person}{Karan Singh}, {and} \bibinfo{person}{Adrien Bousseau}.} \bibinfo{year}{2022}\natexlab{}.
\newblock \showarticletitle{Piecewise-smooth surface fitting onto unstructured 3D sketches}.
\newblock \bibinfo{journal}{\emph{ACM Transactions on Graphics (TOG)}} \bibinfo{volume}{41}, \bibinfo{number}{4} (\bibinfo{year}{2022}), \bibinfo{pages}{1--16}.
\newblock


\bibitem[Zhang et~al\mbox{.}(2022)]%
        {zhang2022implicit}
\bibfield{author}{\bibinfo{person}{Congyi Zhang}, \bibinfo{person}{Mohamed Elgharib}, \bibinfo{person}{Gereon Fox}, \bibinfo{person}{Min Gu}, \bibinfo{person}{Christian Theobalt}, {and} \bibinfo{person}{Wenping Wang}.} \bibinfo{year}{2022}\natexlab{}.
\newblock \showarticletitle{An Implicit Parametric Morphable Dental Model}.
\newblock \bibinfo{journal}{\emph{ACM Transactions on Graphics (TOG)}} \bibinfo{volume}{41}, \bibinfo{number}{6} (\bibinfo{year}{2022}), \bibinfo{pages}{1--13}.
\newblock


\bibitem[Zhu et~al\mbox{.}(2020)]%
        {zhu2020deep}
\bibfield{author}{\bibinfo{person}{Heming Zhu}, \bibinfo{person}{Yu Cao}, \bibinfo{person}{Hang Jin}, \bibinfo{person}{Weikai Chen}, \bibinfo{person}{Dong Du}, \bibinfo{person}{Zhangye Wang}, \bibinfo{person}{Shuguang Cui}, {and} \bibinfo{person}{Xiaoguang Han}.} \bibinfo{year}{2020}\natexlab{}.
\newblock \showarticletitle{Deep Fashion3D: A dataset and benchmark for 3D garment reconstruction from single images}. In \bibinfo{booktitle}{\emph{European Conference on Computer Vision}}. Springer, \bibinfo{pages}{512--530}.
\newblock


\end{thebibliography}

\end{document}